\documentclass[12pt]{iopart}
\usepackage{iopams}

\usepackage{graphicx}
\usepackage{color}
\usepackage{amsfonts}

\begin{document}

\title[]{Hadron transverse momentum distributions in the Tsallis statistics with escort probabilities}

\author{A.S.~Parvan$^{1,2}$}

\address{$^1$ Bogoliubov Laboratory of Theoretical Physics, Joint Institute for Nuclear Research, Dubna, Russia}
\address{$^2$ Department of Theoretical Physics, Horia Hulubei National Institute for R$\&$D in Physics and Nuclear Engineering, Bucharest-Magurele, Romania}

\eads{\mailto{parvan@theor.jinr.ru}, \mailto{parvan@theory.nipne.ro}}
%\ead{parvan@theory.nipne.ro}
%\vspace{10pt}
%\begin{indented}
%\item[]August 2022
%\end{indented}

\begin{abstract}
The exact and approximate hadron transverse momentum distributions for the Fermi-Dirac, Bose-Einstein and Maxwell-Boltzmann statistics of particles in the framework of the Tsallis statistics with escort probabilities (the Tsallis-3 statistics) have been derived. The classical and quantum transverse momentum distributions in the zeroth term approximation and the quantum transverse momentum distributions in the factorization approximation introduced in the zeroth term approximation were found. The transverse momentum distributions in the zeroth term approximation and in the factorization approximation of the zeroth term approximation are the same in the Tsallis-3, Tsallis-2 and $q$-dual statistics. The well-known classical phenomenological Tsallis distribution exactly coincides with the classical transverse momentum distribution of the Tsallis-3 statistics in the zeroth term approximation for which the entropy of system is zero in the whole range of state variables. However, the quantum phenomenological Tsallis distribution does not coincide with either the exact or approximate transverse momentum distributions of the Tsallis-3 statistics. The exact Tsallis-3 classical distribution and the classical phenomenological Tsallis distribution were applied to describe the experimental spectra of the charged pions produced in the proton-proton collisions at high energies. The values of the parameters $(T,q)$ for both these model distributions differ in the whole energy range. Thus, the classical phenomenological Tsallis distribution is an unsatisfactory approximation for the exact classical transverse momentum distribution of the Tsallis-3 statistics.
\end{abstract}

%\maketitle

\section{Introduction}\label{sec1}
Nowadays, the Tsallis statistics~\cite{Tsal88,Tsal98} is successfully used in different fields of physics. However, in high-energy physics, the Tsallis-like distributions~\cite{Cleymans09,Alberico09,Wong15,Shen18,PHENIX1,CMS1,CMS2} and the phenomenological Tsallis distributions~\cite{Cleymans12a,Cleymans2012,Bediaga00,Beck00} are applied instead of the exact distributions of the Tsallis statistics. These simple functions are usually presented as the distributions of the Tsallis statistics and successfully  applied to describe the experimental spectra of hadrons produced in relativistic heavy-ion collisions and proton-proton reactions at LHC and RHIC energies~\cite{Alberico09,Wong15,PHENIX1,CMS1,CMS2,Cleymans12a,Cleymans2012,Rybczynski14,Cleymans13,Azmi14,Marques13,Li14,Zhang22,Azmi20,Biro20,Che21,Parvan17a,Waqas20,Yang21,Wibig14,Azmi15,Gu22,Chen21,Aaij21,Kyan22,Wong13,Zheng15,Prosz15,Grigoryan17,Wilk13,Tao21,Biyajima06,Marques15,Shen2019a,Chaturvedi2018}.
However, their belonging to the Tsallis statistics is questionable and has not been proven.

The Tsallis-like transverse momentum  distributions~\cite{Alberico09,Wong15,Shen18,PHENIX1,CMS1,CMS2,Zhang22,Yang21,Wibig14,Aaij21,Wong13,Prosz15,Biyajima06,Shen2019a} (see~\ref{App1})
\begin{equation}\label{i1}
\varepsilon_{\mathbf{p}} \frac{d^{3}N}{d^{3}p} \equiv A \sum\limits_{\sigma}\langle n_{\mathbf{p}\sigma} \rangle = g A  \frac{1}{\left[1- (1-q) \frac{\varepsilon_{\mathbf{p}}-\mu}{T} \right]^{\frac{1}{q-1}}+\eta}
\end{equation}
and the phenomenological Tsallis transverse momentum distributions~\cite{Cleymans12a,Cleymans2012,Rybczynski14,Cleymans13,Azmi14,Marques13,Li14,Azmi20,Biro20,Parvan17a,Azmi15,Kyan22,Zheng15,Tao21,Marques15,Chaturvedi2018}
\begin{equation}\label{i2}
\varepsilon_{\mathbf{p}} \frac{d^{3}N}{d^{3}p} \equiv \frac{V}{(2\pi)^{3}} \varepsilon_{\mathbf{p}} \sum\limits_{\sigma}  \langle n_{\mathbf{p}\sigma} \rangle^{q} = \frac{g V}{(2\pi)^{3}} \varepsilon_{\mathbf{p}}  \left( \frac{1}{\left[1- (1-q) \frac{\varepsilon_{\mathbf{p}}-\mu}{T} \right]^{\frac{1}{q-1}}+\eta}\right)^{q}
\end{equation}
are based on the Fermi-Dirac ($\eta=1$), Bose-Einstein ($\eta=-1$) and Maxwell-Boltzmann ($\eta=0$) single-particle distribution functions (mean occupation numbers) of the Tsallis statistics~\cite{Buyukkilic}:
\begin{eqnarray}\label{i3}
  \langle n_{\mathbf{p}\sigma} \rangle &=& \left[1-(1-q)\frac{\varepsilon_{\mathbf{p}}-\mu}{T}\right]^{\frac{1}{1-q}} \qquad \;  \mathrm{for} \quad \eta =0, \\ \label{i4}
  \langle n_{\mathbf{p}\sigma} \rangle &=& \frac{1}{\left[1-(1-q)\frac{\varepsilon_{\mathbf{p}}-\mu}{T}\right]^{\frac{1}{q-1}}\pm 1} \qquad  \mathrm{for} \quad \eta = \pm 1,
\end{eqnarray}
where $d^{3}p=\varepsilon_{\mathbf{p}} p_{T}dp_{T}dyd\varphi$, $\varepsilon_{\mathbf{p}}=m_{T} \cosh y$ is the single-particle energy, $m_{T}=\sqrt{p_{T}^{2}+m^{2}}$ is the transverse mass, $p_{T},y$ and $\varphi$ are the transverse momentum, rapidity and azimuthal angle, respectively, $V$ is the volume, $T$ is the temperature, $\mu$ is the chemical potential and $g$ is the spin degeneracy factor. The classical and quantum Tsallis-like transverse momentum  distributions (\ref{i1}) can be explicitly rewritten as~\cite{Alberico09,Wong15,Shen18,PHENIX1,CMS1,CMS2,Zhang22,Yang21,Wibig14,Aaij21,Wong13,Prosz15,Biyajima06,Shen2019a}
\begin{eqnarray}\label{i5}
\frac{d^{2}N}{dp_{T}dy} &=&  2\pi p_{T} g A  \left[1-(1-q) \frac{m_{T} \cosh y-\mu}{T} \right]^{\frac{1}{1-q}} \qquad  \mathrm{for} \quad \eta =0, \\ \label{i5a}
\frac{d^{2}N}{dp_{T}dy} &=&  2\pi p_{T} g A  \frac{1}{\left[1- (1-q) \frac{m_{T} \cosh-\mu}{T} \right]^{\frac{1}{q-1}}\pm 1} \qquad \;\;\; \mathrm{for} \quad \eta = \pm 1.
\end{eqnarray}
The classical and quantum phenomenological Tsallis transverse momentum distributions (\ref{i2}) can be explicitly rewritten as~\cite{Cleymans12a,Cleymans2012,Rybczynski14,Cleymans13,Azmi14,Marques13,Li14,Azmi20,Biro20,Parvan17a,Azmi15,Kyan22,Zheng15,Tao21,Marques15,Chaturvedi2018}
\begin{eqnarray}\label{i6}
\frac{d^{2}N}{dp_{T}dy} &=& \frac{gV}{(2\pi)^{2}} p_{T}  m_{T} \cosh y  \left[1-(1-q) \frac{m_{T} \cosh y-\mu}{T} \right]^{\frac{q}{1-q}} \;\;  \mathrm{for} \; \eta =0, \\ \label{i6a}
\frac{d^{2}N}{dp_{T}dy} &=& \frac{gV}{(2\pi)^{2}} \frac{p_{T}  m_{T} \cosh y}{\left(\left[1- (1-q) \frac{m_{T} \cosh - \mu}{T} \right]^{\frac{1}{q-1}}\pm 1\right)^{q}} \qquad  \qquad \mathrm{for} \; \eta = \pm 1.
\end{eqnarray}
In accordance with the requirements of the relativistic statistical mechanics the multiplier $A$ in Eqs.~(\ref{i1}), (\ref{i5}), (\ref{i5a}) should be $A=V \varepsilon_{\mathbf{p}}/(2\pi)^{3}$ (see~\ref{App1}), however, its form in Refs.~\cite{Wong15,Shen18,PHENIX1,CMS1,CMS2,Zhang22,Yang21,Wibig14,Aaij21,Wong13,Prosz15,Biyajima06,Shen2019a} differs essentially except of Ref.~\cite{Alberico09}. The Tsallis blast-wave models~\cite{Che21,Waqas20,Gu22,Chen21,Grigoryan17} are also constructed on the base of Eqs.~(\ref{i3}) and (\ref{i4}) given in the form $\langle n_{\mathbf{p}\sigma} \rangle$ or $\langle n_{\mathbf{p}\sigma} \rangle^{q}$.

The classical and quantum single-particle distribution functions of the Tsallis statistics~(\ref{i3}) and (\ref{i4}) obtained in Ref.~\cite{Buyukkilic} are inconsistent (see the proof in Ref.~\cite{Parvan2021a}). First, these distributions were erroneously calculated in Ref.~\cite{Buyukkilic} from the first principles of the Tsallis statistics. Second, they were derived in the framework of the Tsallis-2 statistics, which is generally inconsistent due to the improper definition of generalized mean values for which $\langle 1\rangle \neq 1$ (see Ref.~\cite{Tsal98}). Third, these distributions were obtained in the factorization approximation which is evidently mathematically invalid. It factorizes a power-law function as if it were an exponential one (for details see Ref.~\cite{Parvan2021a}). Thus, the Tsallis-like distributions (\ref{i1}), (\ref{i5}), (\ref{i5a}), the phenomenological Tsallis distributions (\ref{i2}), (\ref{i6}), (\ref{i6a}) and the Tsallis blast-wave models~\cite{Che21,Waqas20,Gu22,Chen21,Grigoryan17} based on the distributions~(\ref{i3}) and (\ref{i4}) are questionable in the framework of the Tsallis statistics. In Refs.~\cite{Parvan17,Parvan2020a}, it was analytically demonstrated that the classical phenomenological Tsallis distribution~(\ref{i6}) for the Maxwell-Boltzmann statistics of particles exactly corresponds to the transverse momentum distribution in the zeroth term approximation of the Tsallis-2 statistics. Thus, the classical phenomenological Tsallis distribution~(\ref{i6}) in the Tsallis statistics is not founded from the point of view of the fundamentals of the statistical mechanics. However, the classical phenomenological Tsallis distribution~(\ref{i6}) may be consistent in the $q$-dual statistics (see Ref.~\cite{Parvan2020b}). The classical and quantum Tsallis-like distributions~(\ref{i1}), (\ref{i5}), (\ref{i5a}) and the quantum phenomenological Tsallis distributions~(\ref{i2}), (\ref{i6a})~\cite{Cleymans12a,Cleymans2012} for the Fermi-Dirac and Bose-Einstein statistics of particles correspond neither to the exact transverse momentum distributions nor to the zeroth term approximation distributions of the Tsallis-1, Tsallis-2 statistics~\cite{Parvan2020a} and the $q$-dual statistics~\cite{Parvan2020b}. In the present work, we will demonstrate that the classical and quantum Tsallis-like distributions~(\ref{i1}), (\ref{i5}), (\ref{i5a}) and the quantum phenomenological Tsallis distributions~(\ref{i2}), (\ref{i6a}) do not correspond to the distributions of the Tsallis-3 statistics either.

Nowadays, there are at least three versions of the Tsallis statistics~\cite{Tsal88,Tsal98} based on the same generalized entropy~\cite{Wehrl,Havrda,Daroczy} known as the Tsallis entropy~\cite{Tsal88,Tsal98} (see Ref.~\cite{Tsal_book} for more explanations) which differ from each other only in the definition of the mathematical mean values of the operators. The first variant of the Tsallis statistics~\cite{Tsal88,Tsal98}, which is also called the Tsallis-1 statistics, is defined by the standard mean values, as in the Boltzmann-Gibbs statistics. Such mathematical expectation values are consistent with the normalization condition of probabilities in full accordance with the requirements of statistical mechanics and probability theory. The second version of the Tsallis statistics~\cite{Tsal88,Tsal98,Curado91}, which is also called the Tsallis-2 statistics, is defined by the generalized mean values of the operators, which do not agree with the probability normalization condition. Such unconventional mathematical expectation values lead to an inconsistent relationship between statistical mechanics, probability theory and the theory of equilibrium thermodynamics due to the fact that $\langle 1\rangle \neq 1$. The third version of the Tsallis statistics~\cite{Tsal98}, called the Tsallis-3 statistics, uses the normalized generalized mean values of the operators. However, in contrast to the Tsallis-2 statistics, the mean values of the Tsallis-3 statistics are consistent with the normalization condition for the probabilities of microstates of the system. One of the important properties of the Tsallis statistics is the invariance of its probability distribution under the uniform shift of the energy spectrum. The probabilities of microstates of the Tsallis-1 statistics, the Tsallis-3 statistics and the $q$-dual statistics are invariant under such homogeneous energy translations~\cite{Parvan2021c,Sisto}. However, the probability distribution of the Tsallis-2 statistics is not invariant under this transformation~\cite{Tsal98,Parvan2021c}.

For the first time, the exact transverse momentum distributions in the grand canonical ensemble of the Tsallis-1 statistics were calculated in  Refs.~\cite{Parvan17,Parvan16,Parvan17aa}. These distributions were only derived for the Maxwell-Boltzmann statistics of massless particles in the ultrarelativistic approximation. Only in this case the exact analytical expressions for the thermodynamic quantities of the Tsallis-1 statistics can be calculated definitely~\cite{Parvan17,Parvan16,Parvan17aa}. However, in the case of the relativistic massive particles, the exact formulae for the Maxwell-Boltzmann, Fermi-Dirac and Bose-Einstein statistics of particles can be expressed only in the integral representation~\cite{Parvan2020a}. The exact transverse momentum distributions were also calculated in the framework of the $q$-dual statistics~\cite{Parvan2020b}. For the first time, the exact Tsallis-1 transverse momentum distributions were applied to describe the experimental data of hadrons produced in proton-proton collisions at LHC and RHIC energies in Refs.~\cite{Parvan16,Parvan2020a}. In Ref.~\cite{Parvan16}, the numerical results for the exact Tsallis-1 classical distribution and the classical phenomenological Tsallis distribution for the massless particles in the ultrarelativistic approximation were compared. However, in Ref.~\cite{Parvan2020a}, the numerical results for the exact Tsallis-1 classical distribution of the relativistic massive particles were obtained.

The main aim of this paper is to derive the exact Tsallis-3 transverse momentum distribution and apply it to describe the experimental transverse momentum spectra of hadrons created in proton-proton collisions at LHC and RHIC energies. In high-energy physics, this has not been done yet. The present calculations are motivated by the fact that the Tsallis-3 statistics is frequently considered by the scientific community to be the most correct (see, for example, the Ref.~\cite{Kapusta}). However, the Tsallis-1 statistics is not inconsistent either (see the proof in Refs.~\cite{Parvan2021c,Parv2,Parv2a}). Another aim of this study is to derive the transverse momentum distribution in the zeroth term approximation of the Tsallis-3 statistics and compare it with the phenomenological Tsallis distribution.

The organization of the paper is as follows. In Sect.~\ref{sec2}, we define the new representation for the general formalism of the Tsallis-3 statistics. In Sect.~\ref{sec3}, we derive the classical and quantum transverse momentum distributions. The experimental data of hadrons are described in Sect.~\ref{sec4}. In Sect.~\ref{sec5}, we summarize and draw conclusions. In this section and throughout the paper we use the system of natural units $\hbar=c=k_{B}=1$.

\section{New formulation of the Tsallis statistics with escort probabilities in the grand canonical ensemble}\label{sec2}
The Tsallis statistics with escort probabilities which is named also the Tsallis-3 statistics~\cite{Tsal98} is defined by the generalized entropy~\cite{Tsal88,Tsal98}
\begin{equation}\label{1}
    S = \sum\limits_{i} \frac{p_{i}^{q}-p_{i}}{1-q} =\frac{1}{\theta}\sum\limits_{i} p_{i}^{q} S_{i}, \qquad  S_{i} = -\theta \frac{p_{i}^{1-q}-1}{1-q}
\end{equation}
and the generalized mean values of operators~\cite{Tsal98}
\begin{eqnarray}\label{2}
   \langle A \rangle &=& \frac{\sum\limits_{i} p_{i}^{q} A_{i}}{\sum\limits_{i} p_{i}^{q}} = \frac{1}{\theta}\sum\limits_{i} p_{i}^{q} A_{i}, \\ \label{3}
   \theta  &\equiv& \sum\limits_{i} p_{i}^{q}
\end{eqnarray}
with the probabilities $p_{i}$ of the microstates of the system normalized to unity:
\begin{equation} \label{1b}
     \phi \equiv \sum\limits_{i} p_{i} - 1 = 0.
\end{equation}
Here the entropic parameter $q\in\mathbb{R}$ is a real number that takes values in the range $0<q<\infty$. Is should be stressed that in the Boltzmann-Gibbs limit $q\to 1$, the entropy (\ref{1}) resembles the usual Boltzmann-Gibbs entropy, $S=-\sum_{i} p_{i} \ln p_{i}$, and the function $\theta=1$. In this case the Tsallis-3 statistics is reduced to the usual Boltzmann-Gibbs statistics.

In the Tsallis-$3$ statistics, the thermodynamic potential $\Omega$ of the grand canonical ensemble is derived from the fundamental thermodynamic potential $\langle H \rangle$ by the Legendre transform:
\begin{eqnarray}\label{4}
 \Omega &=& \langle H \rangle -TS-\mu \langle N \rangle = \frac{1}{\theta}\sum\limits_{i} p_{i}^{q} \Omega_{i}, \\ \label{4b}
 \Omega_{i} &=& -TS_{i} + E_{i} - \mu N_{i}  = T \theta \frac{p_{i}^{1-q}-1}{1-q} + E_{i}-\mu N_{i}.
\end{eqnarray}
Here the mean energy and the mean number of particles of the system are defined as~\cite{Tsal98}
\begin{eqnarray}\label{4c}
 \langle H \rangle &=&  \frac{1}{\theta} \sum_{i}  p_{i}^{q} E_{i}, \\ \label{4d}
 \langle N \rangle &=&  \frac{1}{\theta} \sum_{i}  p_{i}^{q} N_{i},
\end{eqnarray}
where $E_{i}$ and $N_{i}$ are the energy and the number of particles, respectively, in the $i$-th microscopic state of the system.

In the grand canonical ensemble, the unknown equilibrium probabilities $\{p_{i}\}$ are derived from the principle of thermodynamic equilibrium (the principle of maximum entropy) by the method of the Lagrange multipliers (see e.g. Refs.~\cite{Jaynes2,Parvan2015,Krasnov,Parvan2020b}):
\begin{eqnarray}\label{5}
 \Phi &=& \Omega - \lambda \phi, \\ \label{7}
  \frac{\partial \Phi}{\partial p_{i}} &=& 0,
\end{eqnarray}
where $\lambda$ is the Lagrange multiplier and $\Phi$ is the Lagrange function. Substituting Eqs.~(\ref{4}) and (\ref{1b}) into Eqs.~(\ref{5}), (\ref{7}) and using Eqs.~(\ref{2}), (\ref{3}), we obtain
\begin{equation}\label{8}
 p_{i}^{q-1} = \frac{1}{q}\left[1-(1-q)\frac{\lambda}{T}\right] \left[1+(1-q)\frac{\langle H \rangle - E_{i} - \mu (\langle N \rangle -N_{i})}{T\theta}\right]^{-1}.
\end{equation}
Multiplying Eq.~(\ref{8}) by $p_{i}$ and summing it over $i$, we get
\begin{equation}\label{9}
  \frac{1}{q}\left[1-(1-q)\frac{\lambda}{T}\right] =\theta.
\end{equation}
Substituting Eq.~(\ref{9}) into Eq.~(\ref{8}) and using Eqs.~(\ref{1}), (\ref{1b}), (\ref{3}), we obtain the normalized equilibrium probability for the $i$th microstate of the system for the Tsallis-3 statistics in the grand canonical ensemble as
\begin{equation}\label{10}
  p_{i} = \left[1+(1-q)\frac{\Lambda - E_{i}+\mu N_{i}}{T \theta^{2}}\right]^{\frac{1}{1-q}},
\end{equation}
where
\begin{equation}\label{11}
  \Lambda \equiv - \theta T \frac{\theta - 1}{1-q} + \langle H \rangle - \mu \langle N \rangle  = - \theta T S + \langle H \rangle - \mu \langle N \rangle.
\end{equation}
Substituting Eq.~(\ref{10}) into Eqs.~(\ref{3}), (\ref{1b}), we obtain the system of two norm equations for two unknown variables $\Lambda$ and $\theta$:
\begin{eqnarray}\label{12}
   && \sum\limits_{i} \left[1+(1-q)\frac{\Lambda - E_{i}+\mu N_{i}}{T \theta^{2}}\right]^{\frac{1}{1-q}} =1, \\ \label{13}
  && \sum\limits_{i} \left[1+(1-q)\frac{\Lambda - E_{i}+\mu N_{i}}{T \theta^{2}}\right]^{\frac{q}{1-q}} = \theta.
\end{eqnarray}
Here $\Lambda$ and $\theta$ are the normalization functions, which are found by solving the system of equations (\ref{12}) and (\ref{13}). In the Tsallis-3 statistics, there are two normalization functions while in the Tsallis-1 and Tsallis-2 statistics, there is only one~\cite{Tsal98,Parvan2020a}. Note that in the Boltzmann-Gibbs limit $q\to 1$, the probability distribution (\ref{10}) resembles the probability of the Boltzmann-Gibbs statistics $p_{i}=\exp[(\Lambda-E_{i}+\mu N_{i})/T]$, where $\Lambda=-T\ln Z$ is the thermodynamic potential and $Z=\sum_{i} \exp[-(E_{i} - \mu N_{i})/T]$ is the partition function of the Boltzmann-Gibbs statistics in the grand canonical ensemble. Note that the probability distribution (\ref{10}) for the Tsallis-3 statistics in the grand canonical ensemble can be rewritten in other equivalent forms (see~\ref{App2}).

Substituting Eq.~(\ref{10}) into Eq.~(\ref{2}), we get the generalized mean values of the Tsallis-3 statistics in the grand canonical ensemble as
\begin{equation}\label{14}
   \langle A \rangle = \frac{1}{\theta}\sum\limits_{i} A_{i} \left[1+(1-q)\frac{\Lambda - E_{i}+\mu N_{i}}{T \theta^{2}}\right]^{\frac{q}{1-q}}.
\end{equation}

Using Eqs.~(\ref{1})--(\ref{4b}) and (\ref{10}), we can rewrite the entropy and the thermodynamic potential of the Tsallis-3 statistics in the grand canonical ensemble as
\begin{eqnarray}\label{11b}
  S &=& \frac{\theta - 1}{1-q} = -\frac{1}{T\theta}[\Lambda -\langle H \rangle + \mu \langle N \rangle], \\ \label{16}
  \Omega &=& \frac{\Lambda}{\theta} + (1-\frac{1}{\theta})[\langle H \rangle - \mu \langle N \rangle] = \Lambda + TS (\theta-1)  = \Lambda + T \frac{(\theta-1)^{2}}{1-q}.
\end{eqnarray}

In order to rewrite the quantities (\ref{10}), (\ref{12}), (\ref{13}) and (\ref{14}) in the integral representation, we use the formulae for the Gamma-function in the form~\cite{Abramowitz,Prato}:
\begin{eqnarray}\label{17}
  x^{-y} &=& \frac{1}{\Gamma(y)} \int\limits_{0}^{\infty}  t^{y-1} e^{-tx}  dt, \qquad \qquad    \mathrm{Re}(x)>0, \quad   \mathrm{Re}(y)>0, \\ \label{18}
   x^{y-1} &=& \Gamma(y) \frac{i}{2\pi} \oint\limits_{C} (-t)^{-y} e^{-tx}  dt, \qquad   \mathrm{Re}(x)>0, \quad  |y|<\infty.
\end{eqnarray}
Using Eq.~(\ref{17}) for $q>1$ and Eq.~(\ref{18}) for $q<1$, we derive formulae for the probability of microstates (\ref{10}) in the integral representation as
\begin{equation}\label{19}
  p_{i} = \frac{1}{\Gamma\left(\frac{1}{q-1}\right)} \int\limits_{0}^{\infty} t^{\frac{2-q}{q-1}} e^{-t+\beta'\left(\Lambda-\Omega_{G}\left(\beta'\right)\right)} p_{Gi}\left(\beta'\right) dt \qquad \mathrm{for} \quad q>1
\end{equation}
and
\begin{equation}\label{20}
  p_{i} = \Gamma\left(\frac{2-q}{1-q}\right)  \frac{i}{2\pi} \oint\limits_{C} (-t)^{-\frac{2-q}{1-q}}  e^{-t+\beta'\left(\Lambda-\Omega_{G}\left(\beta'\right)\right)} p_{Gi}\left(\beta'\right) dt \quad \mathrm{for} \quad q<1,
\end{equation}
where
\begin{eqnarray}\label{21}
  p_{Gi}\left(\beta'\right) &=& \frac{1}{Z_{G}\left(\beta'\right)} e^{-\beta'(E_{i}-\mu N_{i})}, \\ \label{22}
  Z_{G}\left(\beta'\right) &=& \sum\limits_{i} e^{-\beta'(E_{i}-\mu N_{i})}, \\ \label{23}
  \Omega_{G}\left(\beta'\right) &=& -\frac{1}{\beta'} \ln Z_{G}\left(\beta'\right)
\end{eqnarray}
and
\begin{equation} \label{24}
 \beta'=\frac{-t(1-q)}{T\theta^{2}}.
\end{equation}
Note that the probability distribution of the Tsallis-3 statistics (\ref{19}) and (\ref{20}) is a function of the probability distribution of the Boltzmann-Gibbs statistics (\ref{21}).

In the integral representation, the norm equation (\ref{12}) takes the following form:
\begin{eqnarray}\label{25}
 1 &=& \frac{1}{\Gamma\left(\frac{1}{q-1}\right)} \int\limits_{0}^{\infty} t^{\frac{2-q}{q-1}} e^{-t+\beta'\left(\Lambda-\Omega_{G}\left(\beta'\right)\right)} dt
  \nonumber \\ &=& \sum\limits_{n=0}^{\infty} \frac{1}{n!\Gamma\left(\frac{1}{q-1}\right)} \int\limits_{0}^{\infty} t^{\frac{2-q}{q-1}} e^{-t+\beta' \Lambda} (-\beta'\Omega_{G}\left(\beta'\right))^{n} dt  \qquad   \mathrm{for} \quad q>1
\end{eqnarray}
and
\begin{eqnarray}\label{26}
 1 &=& \Gamma\left(\frac{2-q}{1-q}\right)  \frac{i}{2\pi} \oint\limits_{C} (-t)^{-\frac{2-q}{1-q}} e^{-t+\beta'\left(\Lambda-\Omega_{G}\left(\beta'\right)\right)} dt \nonumber \\ &=& \sum\limits_{n=0}^{\infty} \frac{\Gamma\left(\frac{2-q}{1-q}\right)}{n!}   \frac{i}{2\pi} \oint\limits_{C} (-t)^{-\frac{2-q}{1-q}} e^{-t+\beta' \Lambda} (-\beta'\Omega_{G}\left(\beta'\right))^{n} dt  \;\;\; \mathrm{for} \; q<1,
\end{eqnarray}
where
\begin{equation}\label{27}
  e^{-\beta'\Omega_{G}\left(\beta'\right)}= \sum\limits_{n=0}^{\infty}  \frac{1}{n!} (-\beta'\Omega_{G}\left(\beta'\right))^{n}.
\end{equation}
In this representation, the norm equation (\ref{13}) can be rewritten as
\begin{eqnarray}\label{28}
 \theta &=& \frac{1}{\Gamma\left(\frac{q}{q-1}\right)} \int\limits_{0}^{\infty} t^{\frac{1}{q-1}} e^{-t+\beta'\left(\Lambda-\Omega_{G}\left(\beta'\right)\right)} dt
  \nonumber \\ &=& \sum\limits_{n=0}^{\infty} \frac{1}{n!\Gamma\left(\frac{q}{q-1}\right)} \int\limits_{0}^{\infty} t^{\frac{1}{q-1}} e^{-t+\beta' \Lambda} (-\beta'\Omega_{G}\left(\beta'\right))^{n} dt  \qquad   \mathrm{for} \quad q>1
\end{eqnarray}
and
\begin{eqnarray}\label{29}
 \theta &=& \Gamma\left(\frac{1}{1-q}\right)  \frac{i}{2\pi} \oint\limits_{C} (-t)^{-\frac{1}{1-q}} e^{-t+\beta'\left(\Lambda-\Omega_{G}\left(\beta'\right)\right)} dt \nonumber \\ &=& \sum\limits_{n=0}^{\infty} \frac{\Gamma\left(\frac{1}{1-q}\right)}{n!}   \frac{i}{2\pi} \oint\limits_{C} (-t)^{-\frac{1}{1-q}} e^{-t+\beta' \Lambda} (-\beta'\Omega_{G}\left(\beta'\right))^{n} dt  \;\;\; \mathrm{for} \; q<1.
\end{eqnarray}
Using Eq.~(\ref{17}) for $q>1$ and Eq.~(\ref{18}) for $q<1$, we rewrite the generalized mean values (\ref{14}) of the Tsallis-3 statistics in the form
\begin{eqnarray}\label{30}
  \langle A \rangle &=&  \frac{1}{\theta\Gamma\left(\frac{q}{q-1}\right)} \int\limits_{0}^{\infty}  t^{\frac{1}{q-1}} e^{-t+\beta'\left(\Lambda-\Omega_{G}\left(\beta'\right)\right)} \langle A \rangle_{G}\left(\beta'\right) dt   \nonumber \\
  &=& \frac{1}{\theta} \sum\limits_{n=0}^{\infty}   \frac{1}{n!\Gamma\left(\frac{q}{q-1}\right)} \int\limits_{0}^{\infty}  t^{\frac{1}{q-1}} e^{-t+\beta' \Lambda} (-\beta'\Omega_{G}\left(\beta'\right))^{n} \langle A \rangle_{G}\left(\beta'\right) dt
  \nonumber \\ &&   \mathrm{for} \quad q>1
\end{eqnarray}
and
\begin{eqnarray}\label{31}
   \langle A \rangle &=& \frac{\Gamma\left(\frac{1}{1-q}\right)}{\theta}   \frac{i}{2\pi}  \oint\limits_{C} (-t)^{-\frac{1}{1-q}} e^{-t+\beta'\left(\Lambda-\Omega_{G}\left(\beta'\right)\right)} \langle A \rangle_{G}\left(\beta'\right) dt \nonumber \\
    &=& \frac{1}{\theta} \sum\limits_{n=0}^{\infty} \frac{\Gamma\left(\frac{1}{1-q}\right)}{n!} \frac{i}{2\pi}  \oint\limits_{C} (-t)^{-\frac{1}{1-q}} e^{-t+\beta' \Lambda} (-\beta'\Omega_{G}\left(\beta'\right))^{n} \langle A \rangle_{G}\left(\beta'\right) dt
    \nonumber \\ &&  \mathrm{for}\quad q<1,
\end{eqnarray}
where~\cite{Parvan2020b}
\begin{equation}\label{32}
  \langle A \rangle_{G}\left(\beta'\right) =\frac{1}{Z_{G}\left(\beta'\right)} \sum\limits_{i} A_{i} e^{-\beta'(E_{i}-\mu N_{i})}.
\end{equation}
Note that the generalized mean value (\ref{30}) and (\ref{31}) of the Tsallis-3 statistics is a function of the corresponding mean value (\ref{32}) of the Boltzmann-Gibbs statistics.

\section{Tsallis-3 statistics transverse momentum distribution}\label{sec3}
Let us calculate the transverse momentum distribution of the Tsallis-3 statistics using the relativistic ideal gas of hadrons in the grand canonical ensemble.

\subsection{Exact model}
\subsubsection{Quantum and classical statistics of relativistic particles}
Let us calculate the exact results for the quantities of the relativistic ideal gas of hadrons for the Boze-Einstein $(\eta=-1)$, Fermi-Dirac $(\eta=1)$ and Maxwell-Boltzmann $(\eta=0)$ statistics of particles in the grand canonical ensemble in the framework of the Tsallis-3 statistics.

The thermodynamic potential and the mean occupation numbers for the relativistic ideal gas in the framework of the Boltzmann-Gibbs statistics in the grand canonical ensemble can be written as~\cite{Parvan2020b}
\begin{eqnarray}\label{35}
  -\beta'\Omega_{G}\left(\beta'\right) &=& \sum\limits_{\mathbf{p},\sigma} \ln \left[1+\eta e^{-\beta' (\varepsilon_{\mathbf{p}}-\mu)} \right]^{\frac{1}{\eta}} \qquad   \mathrm{for} \quad \eta=-1,0,1,  \\ \label{36}
  \langle n_{\mathbf{p}\sigma} \rangle_{G}\left(\beta'\right) &=& \frac{1}{e^{\beta' (\varepsilon_{\mathbf{p}}-\mu)}+\eta},
\end{eqnarray}
where $\beta'$ is a function defined in Eq.~(\ref{24}) and $\varepsilon_{\mathbf{p}}=\sqrt{\mathbf{p}^{2}+m^{2}}$ is the relativistic dispersion relation.

The norm functions $\Lambda$ and $\theta$ of the relativistic ideal gas for the Tsallis-3 statistics are calculated from Eqs.~(\ref{25}), (\ref{26}), (\ref{28}), (\ref{29}) and Eq.~(\ref{35}).

The mean occupation numbers of the relativistic ideal gas in the grand canonical ensemble of the Tsallis-3 statistics are calculated using Eqs.~(\ref{30}), (\ref{31}) and (\ref{36}):
\begin{eqnarray}\label{37}
  \langle n_{\mathbf{p}\sigma} \rangle &=&  \frac{1}{\theta\Gamma\left(\frac{q}{q-1}\right)} \int\limits_{0}^{\infty}  t^{\frac{1}{q-1}} e^{-t+\beta'\left(\Lambda-\Omega_{G}\left(\beta'\right)\right)} \frac{1}{e^{\beta' (\varepsilon_{\mathbf{p}}-\mu)}+\eta} dt   \nonumber \\
  &=& \frac{1}{\theta} \sum\limits_{n=0}^{\infty}   \frac{1}{n!\Gamma\left(\frac{q}{q-1}\right)} \int\limits_{0}^{\infty}  t^{\frac{1}{q-1}} e^{-t+\beta' \Lambda} \frac{(-\beta'\Omega_{G}\left(\beta'\right))^{n}}{e^{\beta' (\varepsilon_{\mathbf{p}}-\mu)}+\eta} dt
  \nonumber \\ &&   \mathrm{for} \quad q>1
\end{eqnarray}
and
\begin{eqnarray}\label{38}
   \langle n_{\mathbf{p}\sigma} \rangle &=& \frac{\Gamma\left(\frac{1}{1-q}\right)}{\theta}   \frac{i}{2\pi}  \oint\limits_{C} (-t)^{-\frac{1}{1-q}} e^{-t+\beta'\left(\Lambda-\Omega_{G}\left(\beta'\right)\right)} \frac{1}{e^{\beta' (\varepsilon_{\mathbf{p}}-\mu)}+\eta} dt \nonumber \\
    &=& \frac{1}{\theta} \sum\limits_{n=0}^{\infty} \frac{\Gamma\left(\frac{1}{1-q}\right)}{n!} \frac{i}{2\pi}  \oint\limits_{C} (-t)^{-\frac{1}{1-q}} e^{-t+\beta' \Lambda} \frac{(-\beta'\Omega_{G}\left(\beta'\right))^{n}}{e^{\beta' (\varepsilon_{\mathbf{p}}-\mu)}+\eta} dt
    \nonumber \\ &&  \mathrm{for}\quad q<1.
\end{eqnarray}

It is easy to show that the transverse momentum distribution of particles can be expressed as a function of the mean occupation numbers~\cite{Parvan17,Parvan2020b} (see~\ref{App1}):
\begin{equation}\label{39}
  \frac{d^{2}N}{dp_{T}dy} = \frac{V}{(2\pi)^{3}} \int\limits_{0}^{2\pi} d\varphi p_{T} \varepsilon_{\mathbf{p}} \ \sum\limits_{\sigma} \langle n_{\mathbf{p}\sigma}\rangle,
\end{equation}
where $\varepsilon_{\mathbf{p}}=m_{T} \cosh y$, $m_{T}=\sqrt{p_{T}^{2}+m^{2}}$ is the transverse mass and $p_{T},y$ and $\varphi$ are the transverse momentum, rapidity and the azimuthal angle, respectively. Substituting Eqs.~(\ref{37}) and (\ref{38}) into Eq.~(\ref{39}), we obtain
\begin{eqnarray}\label{40}
  \frac{d^{2}N}{dp_{T}dy} &=& \frac{gV}{(2\pi)^{2}} p_{T}  m_{T} \cosh y  \frac{1}{\theta\Gamma\left(\frac{q}{q-1}\right)} \int\limits_{0}^{\infty}  t^{\frac{1}{q-1}} e^{-t+\beta'\left(\Lambda-\Omega_{G}\left(\beta'\right)\right)} \nonumber \\ &\times&  \frac{1}{e^{\beta' (m_{T} \cosh y-\mu)}+\eta} dt  = \frac{gV}{(2\pi)^{2}} p_{T}  m_{T} \cosh y \frac{1}{\theta} \sum\limits_{n=0}^{\infty}   \frac{1}{n!\Gamma\left(\frac{q}{q-1}\right)} \nonumber \\ &\times&  \int\limits_{0}^{\infty}  t^{\frac{1}{q-1}} e^{-t+\beta' \Lambda} \frac{(-\beta'\Omega_{G}\left(\beta'\right))^{n}}{e^{\beta' (m_{T} \cosh y-\mu)}+\eta} dt  \qquad  \mathrm{for} \quad q>1
\end{eqnarray}
and
\begin{eqnarray}\label{41}
 \frac{d^{2}N}{dp_{T}dy} &=& \frac{gV}{(2\pi)^{2}} p_{T}  m_{T} \cosh y \frac{\Gamma\left(\frac{1}{1-q}\right)}{\theta}   \frac{i}{2\pi}  \oint\limits_{C} (-t)^{-\frac{1}{1-q}} e^{-t+\beta'\left(\Lambda-\Omega_{G}\left(\beta'\right)\right)}  \nonumber \\ &\times&  \frac{1}{e^{\beta' (m_{T} \cosh y-\mu)}+\eta} dt = \frac{gV}{(2\pi)^{2}} p_{T}  m_{T} \cosh y  \frac{1}{\theta} \sum\limits_{n=0}^{\infty} \frac{\Gamma\left(\frac{1}{1-q}\right)}{n!} \nonumber \\ &\times&  \frac{i}{2\pi}  \oint\limits_{C} (-t)^{-\frac{1}{1-q}} e^{-t+\beta' \Lambda} \frac{(-\beta'\Omega_{G}\left(\beta'\right))^{n}}{e^{\beta' (m_{T} \cosh y-\mu)}+\eta} dt
    \qquad \mathrm{for}\quad q<1.
\end{eqnarray}
The transverse momentum distribution (\ref{40}), (\ref{41}) of the Tsallis-3 statistics differ essentially from the transverse momentum distributions of the Tsallis-1 and Tsallis-2 statistics given in Ref.~\cite{Parvan2020a} due to the two norm functions $\Lambda$ and $\theta$, which are solutions of the system of two equations given by Eqs.~(\ref{25}), (\ref{26}), (\ref{28}), (\ref{29}), and the function $\beta'$, which is a function of the norm function $\theta$ (see Eq.~(\ref{24})). Note that the Tsallis-like distributions (\ref{i1}), (\ref{i5}), (\ref{i5a}) and the phenomenological Tsallis distributions (\ref{i2}), (\ref{i6}), (\ref{i6a}) do not correspond to the exact transverse momentum distributions (\ref{40}), (\ref{41}) of the Tsallis-3 statistics.

\subsubsection{Maxwell-Boltzmann statistics of relativistic particles}
The formulae for the Maxwell-Boltzmann statistics of particles can be explicitly written. Taking the limit $\eta\to 0$ in Eq.~(\ref{35}) and integrating over the 3-dimensional momentum $\mathbf{p}$, we obtain~\cite{Parvan2020b}
\begin{equation}\label{42}
  \Omega_{G}\left(\beta'\right)= - \frac{gV}{2\pi^{2}} \frac{m^{2}}{\beta'^{2}} e^{\beta'\mu} K_{2}\left(\beta'm\right),
\end{equation}
where $\beta'$ is a function defined in Eq.~(\ref{24}) and $K_{\nu}(z)$ is the modified Bessel function of the second kind. Substituting Eq.~(\ref{42}) into Eqs.~(\ref{25}) and (\ref{26}), we get
\begin{eqnarray}\label{43}
 1 &=& \frac{1}{\Gamma\left(\frac{1}{q-1}\right)} \int\limits_{0}^{\infty} t^{\frac{2-q}{q-1}} e^{-t+\beta'\Lambda+\omega t^{-1} e^{\beta' \mu} K_{2}\left(\beta' m \right)} dt
  \nonumber \\ &=& \sum\limits_{n=0}^{\infty} \frac{\omega^{n}}{n!\Gamma\left(\frac{1}{q-1}\right)} \int\limits_{0}^{\infty} t^{\frac{2-q}{q-1}-n} e^{-t+\beta' (\Lambda + \mu n)} (K_{2}\left(\beta' m \right))^{n} dt  \quad   \mathrm{for} \quad q>1
\end{eqnarray}
and
\begin{eqnarray}\label{44}
 1 &=& \Gamma\left(\frac{2-q}{1-q}\right)  \frac{i}{2\pi} \oint\limits_{C} (-t)^{-\frac{2-q}{1-q}} e^{-t+\beta'\Lambda+\omega t^{-1} e^{\beta' \mu} K_{2}\left(\beta' m \right)} dt =  \sum\limits_{n=0}^{\infty} \frac{(-\omega)^{n}}{n!}  \nonumber \\ &\times& \Gamma\left(\frac{2-q}{1-q}\right) \frac{i}{2\pi} \oint\limits_{C} (-t)^{-\frac{2-q}{1-q}-n} e^{-t+\beta' (\Lambda + \mu n)} (K_{2}\left(\beta' m \right))^{n} dt  \;\; \mathrm{for} \; q<1,
\end{eqnarray}
where
\begin{equation}\label{45}
  \omega=\frac{gV}{2\pi^{2}} \frac{m^{2} T\theta^{2}}{q-1}.
\end{equation}
Substituting Eq.~(\ref{42}) into Eqs.~(\ref{28}) and (\ref{29}), we have
\begin{eqnarray}\label{46}
 \theta &=& \frac{1}{\Gamma\left(\frac{q}{q-1}\right)} \int\limits_{0}^{\infty} t^{\frac{1}{q-1}} e^{-t+\beta'\Lambda+\omega t^{-1} e^{\beta' \mu} K_{2}\left(\beta' m \right)} dt
  \nonumber \\ &=& \sum\limits_{n=0}^{\infty} \frac{\omega^{n}}{n!\Gamma\left(\frac{q}{q-1}\right)} \int\limits_{0}^{\infty} t^{\frac{1}{q-1}-n} e^{-t+\beta' (\Lambda + \mu n)} (K_{2}\left(\beta' m \right))^{n} dt  \quad   \mathrm{for} \;\; q>1
\end{eqnarray}
and
\begin{eqnarray}\label{47}
 \theta &=& \Gamma\left(\frac{1}{1-q}\right)  \frac{i}{2\pi} \oint\limits_{C} (-t)^{-\frac{1}{1-q}} e^{-t+\beta'\Lambda+\omega t^{-1} e^{\beta' \mu} K_{2}\left(\beta' m \right)} dt =  \sum\limits_{n=0}^{\infty} \frac{(-\omega)^{n}}{n!}  \nonumber \\ &\times& \Gamma\left(\frac{1}{1-q}\right) \frac{i}{2\pi} \oint\limits_{C} (-t)^{-\frac{1}{1-q}-n} e^{-t+\beta' (\Lambda + \mu n)} (K_{2}\left(\beta' m \right))^{n} dt  \;\; \mathrm{for} \; q<1.
\end{eqnarray}

Substituting Eq.~(\ref{42}) into Eqs.~(\ref{37}) and (\ref{38}) and taking the limit $\eta=0$, we can write
\begin{eqnarray}\label{48}
 \langle n_{\mathbf{p}\sigma} \rangle &=&  \frac{1}{\theta\Gamma\left(\frac{q}{q-1}\right)} \int\limits_{0}^{\infty} t^{\frac{1}{q-1}} e^{-t+\beta'(\Lambda-\varepsilon_{\mathbf{p}}+\mu) +\omega t^{-1} e^{\beta' \mu} K_{2}\left(\beta' m \right)} dt     \nonumber  \\
 &=& \frac{1}{\theta} \sum\limits_{n=0}^{\infty} \frac{\omega^{n}}{n!\Gamma\left(\frac{q}{q-1}\right)}  \int\limits_{0}^{\infty} t^{\frac{1}{q-1}-n} e^{-t+\beta'(\Lambda-\varepsilon_{\mathbf{p}}+\mu(n+1))} (K_{2}\left(\beta' m \right) )^{n} dt \nonumber  \\ && \mathrm{for} \quad q>1
\end{eqnarray}
and
\begin{eqnarray}\label{49}
 \langle n_{\mathbf{p}\sigma} \rangle &=& \frac{\Gamma\left(\frac{1}{1-q}\right)}{\theta}   \frac{i}{2\pi} \oint\limits_{C} (-t)^{-\frac{1}{1-q}} e^{-t+\beta'(\Lambda-\varepsilon_{\mathbf{p}}+\mu) +\omega t^{-1} e^{\beta' \mu} K_{2}\left(\beta' m \right)} dt  \nonumber \\
  &=& \frac{1}{\theta} \sum\limits_{n=0}^{\infty} \frac{(-\omega)^{n}}{n!}  \Gamma\left(\frac{1}{1-q}\right)  \frac{i}{2\pi} \oint\limits_{C} (-t)^{-\frac{1}{1-q}-n} e^{-t+\beta'(\Lambda-\varepsilon_{\mathbf{p}}+\mu(n+1))} \nonumber \\ &\times& (K_{2}\left(\beta' m \right) )^{n} dt \qquad \mathrm{for} \quad q<1.
\end{eqnarray}

Substituting Eq.~(\ref{42}) into Eqs.~(\ref{40}) and (\ref{41}) and taking the limit $\eta=0$, we obtain
\begin{eqnarray}\label{50}
  \frac{d^{2}N}{dp_{T}dy} &=& \frac{gV}{(2\pi)^{2}} p_{T}  m_{T} \cosh y \frac{1}{\theta\Gamma\left(\frac{q}{q-1}\right)} \nonumber \\ &\times&  \int\limits_{0}^{\infty} t^{\frac{1}{q-1}} e^{-t+\beta'(\Lambda-m_{T} \cosh y+\mu) +\omega t^{-1} e^{\beta' \mu} K_{2}\left(\beta' m \right)} dt \nonumber \\ &=&
  \frac{gV}{(2\pi)^{2}} p_{T}  m_{T} \cosh y \frac{1}{\theta} \sum\limits_{n=0}^{\infty} \frac{\omega^{n}}{n!\Gamma\left(\frac{q}{q-1}\right)}  \int\limits_{0}^{\infty} t^{\frac{1}{q-1}-n} \nonumber \\ &\times&  e^{-t+\beta'(\Lambda-m_{T} \cosh y+\mu(n+1))}   (K_{2}\left(\beta' m \right) )^{n} dt \qquad  \mathrm{for} \quad q>1
\end{eqnarray}
and
\begin{eqnarray}\label{51}
 \frac{d^{2}N}{dp_{T}dy} &=& \frac{gV}{(2\pi)^{2}} p_{T}  m_{T} \cosh y \frac{\Gamma\left(\frac{1}{1-q}\right)}{\theta}   \frac{i}{2\pi} \nonumber \\ &\times& \oint\limits_{C} (-t)^{-\frac{1}{1-q}} e^{-t+\beta'(\Lambda-m_{T} \cosh y+\mu) +\omega t^{-1} e^{\beta' \mu} K_{2}\left(\beta' m \right)} dt  \nonumber \\
  &=& \frac{gV}{(2\pi)^{2}} p_{T}  m_{T} \cosh y \frac{1}{\theta} \sum\limits_{n=0}^{\infty} \frac{(-\omega)^{n}}{n!}  \Gamma\left(\frac{1}{1-q}\right)  \frac{i}{2\pi}
   \oint\limits_{C} (-t)^{-\frac{1}{1-q}-n} \nonumber \\ &\times&  e^{-t+\beta'(\Lambda-m_{T} \cosh y+\mu(n+1))} (K_{2}\left(\beta' m \right) )^{n} dt \qquad \mathrm{for} \quad q<1.
\end{eqnarray}
The Maxwell-Boltzmann transverse momentum distribution (\ref{50}), (\ref{51}) of the Tsallis-3 statistics differ essentially from the Maxwell-Boltzmann transverse momentum distributions of the Tsallis-1 and Tsallis-2 statistics given in Ref.~\cite{Parvan2020a} due to the two norm functions $\Lambda$ and $\theta$, which are solutions of the system of two equations given by Eqs.~(\ref{43}), (\ref{44}), (\ref{46}), (\ref{47}), and the functions $\beta'$ and $\omega$, which are the functions of the norm function $\theta$ (see Eqs.~(\ref{24}) and (\ref{45})). Note that the classical Tsallis-like distribution (\ref{i5}) and the classical phenomenological Tsallis distribution (\ref{i6}) do not correspond to the exact Maxwell-Boltzmann transverse momentum distributions (\ref{50}), (\ref{51}) of the Tsallis-3 statistics.

\subsubsection{Maxwell-Boltzmann statistics of ultrarelativistic particles}
Let us consider the ideal gas of the Maxwell-Boltzmann massless particles in the ultrarelativistic approximation in the Tsallis-3 statistics. In the ultrarelativistic limit $m\to 0$, the thermodynamic potential (\ref{42}) takes the form~\cite{Parvan17}
\begin{equation}\label{52}
  \Omega_{G}\left(\beta'\right)= - \frac{gV}{\pi^{2}} \frac{1}{\beta'^{4}} e^{\beta'\mu},
\end{equation}
where $\beta'$ is a function given in Eq.~(\ref{24}). Substituting Eq.~(\ref{52}) into Eqs.~(\ref{25}) and (\ref{26}) and using Eqs.~(\ref{17}), (\ref{18}), we obtain
\begin{eqnarray}\label{53}
 1 &=& \frac{1}{\Gamma\left(\frac{1}{q-1}\right)} \int\limits_{0}^{\infty} t^{\frac{2-q}{q-1}} e^{-t+\beta'\Lambda+\tilde{\omega} t^{-3} (q-1)^{-3} e^{\beta' \mu}} dt
  \nonumber \\ &=& \sum\limits_{n=0}^{\infty} \frac{\tilde{\omega}^{n}}{n!}\frac{\Gamma\left(\frac{1}{q-1}-3n\right)}{(q-1)^{3n}\Gamma\left(\frac{1}{q-1}\right)}
  \left[1+(1-q)\frac{\Lambda + \mu n}{T\theta^{2}} \right]^{\frac{1}{1-q}+3n}  \quad   \mathrm{for} \; q>1
\end{eqnarray}
and
\begin{eqnarray}\label{54}
 1 &=& \Gamma\left(\frac{2-q}{1-q}\right)  \frac{i}{2\pi} \oint\limits_{C} (-t)^{-\frac{2-q}{1-q}} e^{-t+\beta'\Lambda+\tilde{\omega} t^{-3} (q-1)^{-3} e^{\beta' \mu}} dt = \sum\limits_{n=0}^{\infty} \frac{\tilde{\omega}^{n}}{n!} \nonumber \\ &\times&   \frac{\Gamma\left(\frac{2-q}{1-q}\right)}{(1-q)^{3n}\Gamma\left(\frac{2-q}{1-q}+3n\right)}
  \left[1+(1-q)\frac{\Lambda + \mu n}{T\theta^{2}} \right]^{\frac{1}{1-q}+3n}  \quad \mathrm{for} \; q<1,
\end{eqnarray}
where
\begin{equation}\label{55}
  \tilde{\omega}=\frac{gV}{\pi^{2}} T^{3}\theta^{6}.
\end{equation}
Substituting Eq.~(\ref{52}) into Eqs.~(\ref{28}) and (\ref{29}) and using Eqs.~(\ref{17}), (\ref{18}), we have
\begin{eqnarray}\label{56}
 \theta &=& \frac{1}{\Gamma\left(\frac{q}{q-1}\right)} \int\limits_{0}^{\infty} t^{\frac{1}{q-1}} e^{-t+\beta'\Lambda+\tilde{\omega} t^{-3} (q-1)^{-3} e^{\beta' \mu}} dt
  \nonumber \\ &=& \sum\limits_{n=0}^{\infty} \frac{\tilde{\omega}^{n}}{n!}\frac{\Gamma\left(\frac{q}{q-1}-3n\right)}{(q-1)^{3n}\Gamma\left(\frac{q}{q-1}\right)}
  \left[1+(1-q)\frac{\Lambda + \mu n}{T\theta^{2}} \right]^{\frac{q}{1-q}+3n}  \quad   \mathrm{for} \; q>1
\end{eqnarray}
and
\begin{eqnarray}\label{57}
  \theta &=& \Gamma\left(\frac{1}{1-q}\right)  \frac{i}{2\pi} \oint\limits_{C} (-t)^{-\frac{1}{1-q}} e^{-t+\beta'\Lambda+\tilde{\omega} t^{-3} (q-1)^{-3} e^{\beta' \mu}} dt = \sum\limits_{n=0}^{\infty} \frac{\tilde{\omega}^{n}}{n!} \nonumber \\ &\times&   \frac{\Gamma\left(\frac{1}{1-q}\right)}{(1-q)^{3n}\Gamma\left(\frac{1}{1-q}+3n\right)}
  \left[1+(1-q)\frac{\Lambda + \mu n}{T\theta^{2}} \right]^{\frac{q}{1-q}+3n}  \quad \mathrm{for} \; q<1.
\end{eqnarray}

Substituting Eq.~(\ref{52}) into Eqs.~(\ref{37}) and (\ref{38}) and taking the limit $\eta=0$, we get
\begin{eqnarray}\label{58}
 \langle n_{\mathbf{p}\sigma} \rangle &=& \frac{1}{\theta\Gamma\left(\frac{q}{q-1}\right)} \int\limits_{0}^{\infty} t^{\frac{1}{q-1}} e^{-t+\beta'(\Lambda-\varepsilon_{\mathbf{p}}+\mu)+\tilde{\omega} t^{-3} (q-1)^{-3} e^{\beta' \mu}} dt
  \nonumber \\ &=& \frac{1}{\theta} \sum\limits_{n=0}^{\infty} \frac{\tilde{\omega}^{n}}{n!}\frac{\Gamma\left(\frac{q}{q-1}-3n\right)}{(q-1)^{3n}\Gamma\left(\frac{q}{q-1}\right)}
  \nonumber \\ &\times& \left[1+(1-q)\frac{\Lambda-\varepsilon_{\mathbf{p}} + \mu (n+1)}{T\theta^{2}} \right]^{\frac{q}{1-q}+3n} \qquad  \mathrm{for} \quad  q>1
\end{eqnarray}
and
\begin{eqnarray}\label{59}
  \langle n_{\mathbf{p}\sigma} \rangle &=& \frac{\Gamma\left(\frac{1}{1-q}\right)}{\theta} \frac{i}{2\pi} \oint\limits_{C} (-t)^{-\frac{1}{1-q}} e^{-t+\beta'(\Lambda-\varepsilon_{\mathbf{p}}+\mu)+\tilde{\omega} t^{-3} (q-1)^{-3} e^{\beta' \mu}} dt \nonumber \\ &=&  \frac{1}{\theta} \sum\limits_{n=0}^{\infty} \frac{\tilde{\omega}^{n}}{n!} \frac{\Gamma\left(\frac{1}{1-q}\right)}{(1-q)^{3n}\Gamma\left(\frac{1}{1-q}+3n\right)} \nonumber \\ &\times&
  \left[1+(1-q)\frac{\Lambda -\varepsilon_{\mathbf{p}} + \mu (n+1)}{T\theta^{2}} \right]^{\frac{q}{1-q}+3n}  \qquad \mathrm{for} \quad  q<1,
\end{eqnarray}
where $\varepsilon_{\mathbf{p}}=|\mathbf{p}|$. Substituting Eq.~(\ref{52}) into Eqs.~(\ref{40}) and (\ref{41}) and taking the limit $\eta=0$, we have
\begin{eqnarray}\label{60}
  \frac{d^{2}N}{dp_{T}dy} &=& \frac{gV}{(2\pi)^{2}} p_{T}^{2} \cosh y \frac{1}{\theta\Gamma\left(\frac{q}{q-1}\right)} \nonumber \\ &\times& \int\limits_{0}^{\infty} t^{\frac{1}{q-1}} e^{-t+\beta'(\Lambda-p_{T} \cosh y+\mu)+\tilde{\omega} t^{-3} (q-1)^{-3} e^{\beta' \mu}} dt \nonumber \\ &=&
  \frac{gV}{(2\pi)^{2}} p_{T}^{2} \cosh y \frac{1}{\theta} \sum\limits_{n=0}^{\infty} \frac{\tilde{\omega}^{n}}{n!}\frac{\Gamma\left(\frac{q}{q-1}-3n\right)}{(q-1)^{3n}\Gamma\left(\frac{q}{q-1}\right)}
  \nonumber \\ &\times& \left[1+(1-q)\frac{\Lambda-p_{T} \cosh y + \mu (n+1)}{T\theta^{2}} \right]^{\frac{q}{1-q}+3n} \quad  \mathrm{for} \; q>1
\end{eqnarray}
and
\begin{eqnarray}\label{61}
 \frac{d^{2}N}{dp_{T}dy} &=& \frac{gV}{(2\pi)^{2}} p_{T}^{2} \cosh y \frac{\Gamma\left(\frac{1}{1-q}\right)}{\theta} \frac{i}{2\pi} \nonumber \\ &\times& \oint\limits_{C} (-t)^{-\frac{1}{1-q}} e^{-t+\beta'(\Lambda-p_{T} \cosh y+\mu)+\tilde{\omega} t^{-3} (q-1)^{-3} e^{\beta' \mu}} dt  \nonumber \\
  &=& \frac{gV}{(2\pi)^{2}} p_{T}^{2} \cosh y \frac{1}{\theta} \sum\limits_{n=0}^{\infty} \frac{\tilde{\omega}^{n}}{n!} \frac{\Gamma\left(\frac{1}{1-q}\right)}{(1-q)^{3n}\Gamma\left(\frac{1}{1-q}+3n\right)} \nonumber \\ &\times&
  \left[1+(1-q)\frac{\Lambda -p_{T} \cosh y + \mu (n+1)}{T\theta^{2}} \right]^{\frac{q}{1-q}+3n} \quad \mathrm{for} \; q<1,
\end{eqnarray}
where $m=0$ and $m_{T}=p_{T}$. The Maxwell-Boltzmann transverse momentum distribution (\ref{60}), (\ref{61}) in the ultrarelativistic approximation of the Tsallis-3 statistics differ essentially from the Maxwell-Boltzmann transverse momentum distribution in the ultrarelativistic approximation of the Tsallis-1 statistics given in Ref.~\cite{Parvan17} due to the two norm functions $\Lambda$ and $\theta$, which are solutions of the system of two equations given by Eqs.~(\ref{53}), (\ref{54}), (\ref{56}), (\ref{57}), and the function $\tilde{\omega}$, which is a function of the norm function $\theta$ (see Eq.~(\ref{55})).

\subsection{Zeroth term approximation}
Let us consider the relativistic ideal gas of the Tsallis-3 statistics in the zeroth term approximation~\cite{Parvan17,Parvan16}. Taking only the term $n=0$ in Eqs.~(\ref{25}), (\ref{26}), (\ref{28}) and (\ref{29}) and using Eqs.~(\ref{17}) and (\ref{18}), we have
\begin{eqnarray}\label{32a}
  1 &=& \left[1+(1-q)\frac{\Lambda}{T\theta^{2}} \right]^{\frac{1}{1-q}}, \\ \label{32b}
  \theta &=& \left[1+(1-q)\frac{\Lambda}{T\theta^{2}} \right]^{\frac{q}{1-q}}.
\end{eqnarray}
Thus in the zeroth term approximation the norm functions $\Lambda=0$ and $\theta=1$. The entropy of the system (\ref{11b}) takes the value
\begin{equation}\label{32c}
  S=\frac{\theta-1}{1-q} = 0.
\end{equation}
This means that the entropy in the zeroth term approximation is zero for all values of temperature $T$ and chemical potential $\mu$.

Substituting these values of $\Lambda$ and $\theta$ into Eqs.~(\ref{30}) and (\ref{31}) and considering only the zeroth term $n=0$, we get
\begin{equation}\label{33}
 \langle A \rangle =  \frac{1}{\Gamma\left(\frac{q}{q-1}\right)} \int\limits_{0}^{\infty} t^{\frac{1}{q-1}} e^{-t} \langle A \rangle_{G}\left(\beta'\right) dt \qquad \qquad \qquad  \mathrm{for} \quad q>1
\end{equation}
and
\begin{equation}\label{34}
 \langle A \rangle = \Gamma\left(\frac{1}{1-q}\right)  \frac{i}{2\pi} \oint\limits_{C} (-t)^{-\frac{1}{1-q}} e^{-t} \langle A \rangle_{G}\left(\beta'\right) dt \qquad  \mathrm{for} \quad q<1,
\end{equation}
where $\langle A \rangle_{G}\left(\beta'\right)$ is defined in Eq.~(\ref{32}) and $\beta'=-t(1-q)/T$.

Substituting Eq.~(\ref{36}) into Eqs.~(\ref{33}) and (\ref{34}), we obtain
\begin{equation}\label{62}
 \langle n_{\mathbf{p}\sigma} \rangle =  \frac{1}{\Gamma\left(\frac{q}{q-1}\right)} \int\limits_{0}^{\infty}  t^{\frac{1}{q-1}} e^{-t}  \frac{1}{e^{\beta' (\varepsilon_{\mathbf{p}}-\mu)}+\eta} dt  \qquad \qquad \qquad \mathrm{for} \quad q>1
\end{equation}
and
\begin{equation}\label{63}
 \langle n_{\mathbf{p}\sigma} \rangle =  \Gamma\left(\frac{1}{1-q}\right)  \frac{i}{2\pi}  \oint\limits_{C}  (-t)^{-\frac{1}{1-q}} e^{-t}
    \frac{1}{e^{\beta' (\varepsilon_{\mathbf{p}}-\mu)}+\eta} dt  \qquad \mathrm{for} \quad  q<1,
\end{equation}
where $\varepsilon_{\mathbf{p}}=\sqrt{\mathbf{p}^{2}+m^{2}}$. The usual Bose-Einstein $(\eta=-1)$, Fermi-Dirac $(\eta=1)$ and Maxwell-Boltzmann $(\eta=0)$ functions can be combined in the following form:
\begin{eqnarray}\label{64}
\frac{1}{e^{x}+\eta} &=& \sum\limits_{k=0}^{\infty} (-\eta)^{k} e^{-x(k+1)},
\end{eqnarray}
where $|e^{-x}|<1$. Using Eqs.~(\ref{17}), (\ref{18}) and (\ref{62})--(\ref{64}), we obtain (cf. with Ref.~\cite{Parvan2020a})
\begin{equation}\label{65}
\langle n_{\mathbf{p}\sigma} \rangle = \sum\limits_{k=0}^{\infty} (-\eta)^{k} \left[1-(k+1) (1-q) \frac{\varepsilon_{\mathbf{p}}-\mu}{T} \right]^{\frac{q}{1-q}} \quad  \mathrm{for} \;\; \eta=-1,0,1.
\end{equation}
The quantum mean occupation numbers in the zeroth term approximation (\ref{65}) for $1<q<\infty$ can be written as
\begin{equation}\label{65a}
 \langle n_{\mathbf{p}\sigma} \rangle = \left((q-1) \frac{\varepsilon_{\mathbf{p}}-\mu}{T}\right)^{\frac{q}{1-q}} \zeta\left(\frac{q}{q-1},1+\frac{1}{(q-1) \frac{\varepsilon_{\mathbf{p}}-\mu}{T}}\right)  \mathrm{for} \;  \eta=-1,
\end{equation}
\begin{eqnarray}\label{65b}
   \langle n_{\mathbf{p}\sigma} \rangle  &=& \left(2(q-1) \frac{\varepsilon_{\mathbf{p}}-\mu}{T}\right)^{\frac{q}{1-q}} \left[ \zeta\left(\frac{q}{q-1},\frac{1}{2}+\frac{1}{2(q-1) \frac{\varepsilon_{\mathbf{p}}-\mu}{T}}\right) \right. \nonumber \\
   &-& \left. \zeta\left(\frac{q}{q-1},1+\frac{1}{2(q-1) \frac{\varepsilon_{\mathbf{p}}-\mu}{T}}\right) \right] \qquad  \qquad \qquad  \mathrm{for} \;\; \eta=1,
\end{eqnarray}
where $\zeta(s,a)$ is the Hurwitz zeta function for $a\neq 0,-1,-2,\ldots$ and $Re(s)>1$. The classical mean occupation numbers in the zeroth term approximation (\ref{65}) can be rewritten as (cf. with Refs.~\cite{Parvan17,Parvan2020a,Parvan2020b})
\begin{equation}\label{66}
 \langle n_{\mathbf{p}\sigma} \rangle = \left[1-(1-q) \frac{\varepsilon_{\mathbf{p}}-\mu}{T} \right]^{\frac{q}{1-q}} \qquad  \mathrm{for} \quad \eta=0.
\end{equation}

Substituting Eq.~(\ref{65}) and (\ref{66}) into Eq.~(\ref{39}), we obtain the transverse momentum distribution of hadrons for the Fermi-Dirac, Bose-Einstein and Maxwell-Boltzmann statistics of particles in the Tsallis-3 statistics in the zeroth term approximation as (cf. with Refs.~\cite{Parvan2020a,Parvan2020b})
\begin{eqnarray}\label{67}
\frac{d^{2}N}{dp_{T}dy} &=& \frac{gV}{(2\pi)^{2}} p_{T}  m_{T} \cosh y  \sum\limits_{k=0}^{\infty} (-\eta)^{k}  \nonumber \\ && \left[1-(k+1) (1-q) \frac{m_{T} \cosh y-\mu}{T} \right]^{\frac{q}{1-q}} \quad  \mathrm{for} \quad \eta=-1,0,1.
\end{eqnarray}
Note that the transverse momentum distributions (\ref{67}) for the Fermi-Dirac, Bose-Einstein and Maxwell-Boltzmann statistics of particles for the Tsallis-3 statistics in the zeroth term approximation are equivalent to the corresponding transverse momentum distributions for the Tsallis-2 and $q$-dual statistics in the zeroth term approximation (see Refs.~\cite{Parvan2020a,Parvan2020b}). The quantum transverse momentum distributions in the zeroth term approximation (\ref{67}) for $1<q<\infty$ can be written as (cf. Eqs.~(\ref{65a}) and (\ref{65b}))
\begin{eqnarray}\label{67a}
 \frac{d^{2}N}{dp_{T}dy} &=& \frac{gV}{(2\pi)^{2}} p_{T}  m_{T} \cosh y   \left((q-1) \frac{\varepsilon_{\mathbf{p}}-\mu}{T}\right)^{\frac{q}{1-q}}  \nonumber  \\ &\times& \zeta\left(\frac{q}{q-1},1+\frac{1}{(q-1) \frac{\varepsilon_{\mathbf{p}}-\mu}{T}}\right) \qquad  \qquad \mathrm{for} \;\;  \eta=-1,
\end{eqnarray}
\begin{eqnarray}\label{67b}
    \frac{d^{2}N}{dp_{T}dy}   &=& \frac{gV}{(2\pi)^{2}} p_{T}  m_{T} \cosh y  \left(2(q-1) \frac{\varepsilon_{\mathbf{p}}-\mu}{T}\right)^{\frac{q}{1-q}} \nonumber \\
    &\times& \left[\zeta\left(\frac{q}{q-1},\frac{1}{2}+\frac{1}{2(q-1) \frac{\varepsilon_{\mathbf{p}}-\mu}{T}}\right) \right. \nonumber \\
    &-& \left. \zeta\left(\frac{q}{q-1},1+\frac{1}{2(q-1) \frac{\varepsilon_{\mathbf{p}}-\mu}{T}}\right) \right] \qquad  \mathrm{for} \;\; \eta=1.
\end{eqnarray}
Note that the similar expressions for the Tsallis-1 statistics were obtained in Ref.~\cite{Bhatt21}. The classical transverse momentum distribution for the Tsallis-3 statistics in the zeroth term approximation (\ref{67}) can be rewritten as (cf. with Refs.~\cite{Parvan17,Parvan2020a,Parvan2020b})
\begin{equation}\label{68}
\frac{d^{2}N}{dp_{T}dy} = \frac{gV}{(2\pi)^{2}} p_{T}  m_{T} \cosh y  \left[1-(1-q) \frac{m_{T} \cosh y-\mu}{T} \right]^{\frac{q}{1-q}}  \mathrm{for} \; \eta=0.
\end{equation}
Note that the classical transverse momentum distribution (\ref{68}) is the same in the Tsallis-3, Tsallis-2 and $q$-dual statistics (see Refs.~\cite{Parvan2020a,Parvan2020b}). However, it differs from the similar distribution of the Tsallis-1 statistics~\cite{Parvan17,Parvan2020a}. The Maxwell-Boltzmann transverse momentum distribution (\ref{68}) of the Tsallis-3 statistics in the zeroth term approximation exactly coincides with the classical phenomenological Tsallis distribution (\ref{i6}). In the zeroth term approximation the entropy of the Tsallis-3 statistics is zero for all values of temperature $T$ and chemical potential $\mu$ (see Eq.~(\ref{32c})). Thus, in the Tsallis-3 statistics the classical phenomenological Tsallis distribution (\ref{i6}) is questionable. The effective entropy of the ideal gas defined in Refs.~\cite{Cleymans12a,Cleymans2012} for the classical phenomenological Tsallis distribution (\ref{i6}) does not correspond to the entropy (\ref{32c}) of the ideal gas of the Tsallis-3 statistics in the zeroth term approximation. The classical Tsallis-like distribution (\ref{i5}) does not correspond to the classical transverse momentum distribution (\ref{68}) of the Tsallis-3, Tsallis-2 and $q$-dual statistics in the zeroth term approximation~\cite{Parvan2020a,Parvan2020b} and the classical transverse momentum distribution of the Tsallis-1 statistics in the zeroth term approximation given in Refs.~\cite{Parvan17,Parvan2020a}. The quantum phenomenological Tsallis distributions (\ref{i6a}) (for the Fermi-Dirac and Bose-Einstein statistics of particles) introduced in Ref.~\cite{Cleymans12a,Cleymans2012} and the quantum Tsallis-like distributions (\ref{i5a}) are not equivalent to the quantum transverse momentum distributions (\ref{67})--(\ref{67b}) for the Tsallis-3, Tsallis-2 and $q$-dual statistics in the zeroth term approximation and the quantum transverse momentum distributions of the Tsallis-1 statistics in the zeroth term approximation~\cite{Parvan2020a}.

\subsection{Quantum spectra in the factorization approximation of the zeroth term approximation}
Let us consider the factorization approximation adopted in Ref.~\cite{Hasegawa09}, which implies the following mathematically unsanctioned replacement:
\begin{equation}\label{68a}
  \left[1-(k+1) (1-q) \frac{\varepsilon_{\mathbf{p}}-\mu}{T} \right]^{\frac{q}{1-q}} \approx \left[1- (1-q) \frac{\varepsilon_{\mathbf{p}}-\mu}{T} \right]^{\frac{q}{1-q}(k+1)}.
\end{equation}
Substituting Eq.~(\ref{68a}) into Eq.~(\ref{65}) and using equation $\sum_{k=0}^{\infty} (-\eta)^{k} z^{k+1}=(z^{-1}+\eta)^{-1}$ for $|z|<1$, we obtain
\begin{equation}\label{68b}
\langle n_{\mathbf{p}\sigma} \rangle = \frac{1}{\left[1- (1-q) \frac{\varepsilon_{\mathbf{p}}-\mu}{T} \right]^{-\frac{q}{1-q}}\pm 1} \qquad \qquad \mathrm{for} \;\; \eta=\pm 1.
\end{equation}
Similarly, the transverse momentum distribution (\ref{67}) for the Fermi-Dirac and Bose-Einstein statistics of particles for the Tsallis-3 statistics in the factorization approximation of the zeroth term approximation can be rewritten as (cf. with Ref.~\cite{Bhatt21} for the Tsallis-1 statistics)
\begin{equation}\label{68c}
\frac{d^{2}N}{dp_{T}dy} = \frac{gV}{(2\pi)^{2}}   \frac{p_{T}  m_{T} \cosh y}{\left[1- (1-q) \frac{m_{T} \cosh y -\mu}{T} \right]^{\frac{q}{q-1}}\pm 1}  \qquad \mathrm{for} \quad \eta=\pm 1.
\end{equation}
Note that the quantum transverse momentum distribution (\ref{68c}) in the factorization approximation of the zeroth term approximation in the Tsallis-3, Tsallis-2 and $q$-dual statistics is the same because the quantum transverse momentum distribution (\ref{67}) in the zeroth term approximation is the same in all these statistics. However, it differs from the similar quantum distribution of the Tsallis-1 statistics~\cite{Bhatt21}. The quantum Tsallis-like distribution (\ref{i5a}) and the quantum phenomenological Tsallis distribution (\ref{i6a}) do not correspond to the quantum transverse momentum distribution (\ref{68c}) in the factorization approximation of the zeroth term approximation of the Tsallis-3, Tsallis-2 and $q$-dual statistics~\cite{Parvan2020a,Parvan2020b} and the quantum transverse momentum distribution in the factorization approximation of the zeroth term approximation of the Tsallis-1 statistics~\cite{Bhatt21}. Thus, the quantum Tsallis-like distribution (\ref{i5a}) and the quantum phenomenological Tsallis distribution (\ref{i6a}) do not belong to Tsallis-1, Tsallis-2, Tsallis-3 and $q$-dual statistics. The form $([1-(1-q)\frac{m_{T} \cosh y -\mu}{T}]^{\frac{c}{q-1}}\pm 1)^{-1}$, where $c=q,1$, of the quantum distribution in the Tsallis statistics (see, for example, Eqs.~(\ref{i5a}), (\ref{i6a}) and (\ref{68c})) can be only obtained in the factorization approximation~\cite{Buyukkilic,Parvan2021a,Bhatt21}, which is clearly invalid mathematically (see also Eq.~(\ref{68a})). Thus, the quantum Tsallis-like distribution (\ref{i5a}), the quantum phenomenological Tsallis distribution (\ref{i6a}), the quantum transverse momentum distribution (\ref{68c}) of the Tsallis-3, Tsallis-2 and $q$-dual statistics and the similar quantum distribution of the Tsallis-1 statistics given in Ref.~\cite{Bhatt21} are questionable.

\section{Analysis and results}\label{sec4}
Let us apply the Maxwell-Boltzmann transverse momentum distribution of the Tsallis-3 statistics (\ref{50}) for $q>1$ to describe the experimental spectra of hadrons produced in proton-proton collisions at high energies. To obtain the exact Tsallis-3 distribution (\ref{50}), we should solve the system of two norm equations (\ref{43}) and (\ref{46}) with respect to the norm functions $\Lambda$ and $\theta$. Generally, the integrals in Eqs.~(\ref{43}), (\ref{46}) and (\ref{50}) are not convergent due to the divergence of the integrants as $t\to 0$. Physically, this divergence is related to the infinite contribution of the Boltzmann-Gibbs thermodynamic potential (\ref{42}), $\Omega_{G}\left(\beta'\right)\to -\infty$, in Eqs.~(\ref{25}), (\ref{28}), (\ref{30}) at infinite value of temperature, $T'=1/\beta' \to \infty$. However, using the series expansion (\ref{27}), we can rewrite Eqs.~(\ref{43}), (\ref{46}) and (\ref{50}) in the form of infinite series in which integrals of some terms are convergent. To find these terms let us investigate Eqs.~(\ref{43}) and (\ref{46}). In the limit $t\to 0$, the integrants in Eqs.~(\ref{43}) and (\ref{46}) can be written as
\begin{equation}\label{69}
  \varphi_{0}(n,t) \sim t^{\frac{2-q}{q-1}-3n} e^{-t+\beta'(\Lambda +\mu n)} \quad \mathrm{and} \quad    \varphi_{1}(n,t) \sim t^{\frac{1}{q-1}-3n} e^{-t+\beta'(\Lambda +\mu n)}.
\end{equation}
Thus, in the series expansions (\ref{43}) and (\ref{46}) only the terms satisfying the conditions $(2-q)/(q-1)-3n > -1$ and $1/(q-1)-3n > -1$, respectively, are convergent. Then the maximal numbers of finite terms in the series expansions (\ref{43}) and (\ref{46}) can be written as
\begin{equation}\label{70}
  n_{0max}=\left[ \frac{1}{3} \left(\frac{2-q}{q-1}+\delta \right) \right] \quad \mathrm{and} \quad n_{1max}=\left[ \frac{1}{3} \left(\frac{1}{q-1}+\delta \right) \right],
\end{equation}
where $\delta=0.98$. However, not all finite terms are physical. We introduce the single upper cut-off limit of summation in the form
\begin{equation}\label{71}
  n_{0}=\left[ \frac{\nu}{3} \left(\frac{2-q}{q-1}+\delta \right) \right],
\end{equation}
where $0\leqslant \nu \leqslant 1$ is a fixed parameter. This parameter provides the correct Boltzmann-Gibbs limit and an unified description of the system of norm equations and statistical averages. Then the norm equations (\ref{43}) and (\ref{46}) and the transverse momentum distribution (\ref{50}) for $q>1$ can be rewritten as
\begin{eqnarray}\label{72}
 1 &=& \sum\limits_{n=0}^{n_{0}} \frac{\omega^{n}}{n!\Gamma\left(\frac{1}{q-1}\right)} \int\limits_{0}^{\infty} t^{\frac{2-q}{q-1}-n} e^{-t+\beta' (\Lambda + \mu n)} (K_{2}\left(\beta' m \right))^{n} dt, \\ \label{73}
 \theta &=& \sum\limits_{n=0}^{n_{0}} \frac{\omega^{n}}{n!\Gamma\left(\frac{q}{q-1}\right)} \int\limits_{0}^{\infty} t^{\frac{1}{q-1}-n} e^{-t+\beta' (\Lambda + \mu n)} (K_{2}\left(\beta' m \right))^{n} dt
\end{eqnarray}
and
\begin{eqnarray}\label{74}
  \frac{d^{2}N}{dp_{T}dy} &=& \frac{gV}{(2\pi)^{2}} p_{T}  m_{T} \cosh y \frac{1}{\theta} \sum\limits_{n=0}^{n_{0}} \frac{\omega^{n}}{n!\Gamma\left(\frac{q}{q-1}\right)}  \int\limits_{0}^{\infty} t^{\frac{1}{q-1}-n} \nonumber \\ &\times&  e^{-t+\beta'(\Lambda-m_{T} \cosh y+\mu(n+1))}   (K_{2}\left(\beta' m \right) )^{n} dt.
\end{eqnarray}

\begin{figure}[!htb]
\begin{center}
\includegraphics[width=0.8\textwidth]{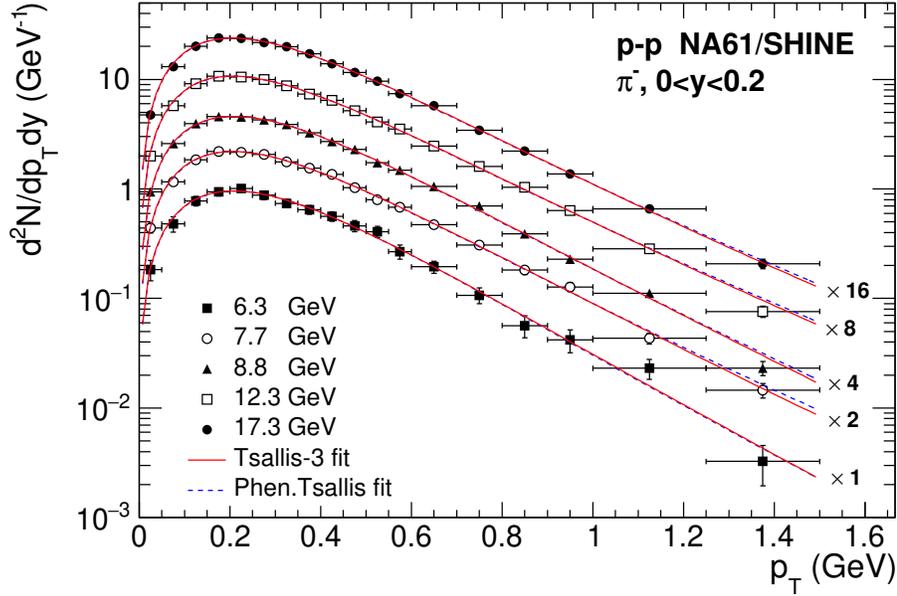} \vspace{-0.3cm}
\caption{(Color online) Transverse momentum spectra of negatively charged pions $\pi^{-}$ measured by the NA61/SHINE Collaboration~\cite{NA61} in $pp$ collisions at $\sqrt{s}=6.3,7.7,8.8,12.3$ and $17.3$ GeV in the rapidity range $0<y<0.2$. The solid and dashed curves are the fits of the data to the exact Tsallis-3 distribution (\ref{75}) and the phenomenological Tsallis distribution (the Tsallis-3 distribution in the zeroth term approximation (\ref{76})), respectively. The symbols represent the experimental data. The numbers next to the lines denote the scaling factor.} \label{fig1}
\end{center}
\end{figure}

The Maxwell-Boltzmann transverse momentum distribution (\ref{74}) of the Tsallis-3 statistics for $q>1$ in the rapidity range $y_{0}\leqslant y \leqslant y_{1}$ takes the form
\begin{eqnarray}\label{75}
 \left. \frac{d^{2}N}{dp_{T}dy}\right|_{y_{0}}^{y_{1}} &=& \frac{gV}{(2\pi)^{2}} p_{T}  m_{T} \int\limits_{y_{0}}^{y_{1}} dy \cosh y \frac{1}{\theta} \sum\limits_{n=0}^{n_{0}} \frac{\omega^{n}}{n!\Gamma\left(\frac{q}{q-1}\right)}  \int\limits_{0}^{\infty} t^{\frac{1}{q-1}-n} \nonumber \\ &\times&  e^{-t+\beta'(\Lambda-m_{T} \cosh y+\mu(n+1))}   (K_{2}\left(\beta' m \right) )^{n} dt.
\end{eqnarray}
For convenience, in further numerical calculations we take $\delta=0$ in Eq.~(\ref{71}).

The Maxwell-Boltzmann transverse momentum distribution of the Tsallis-3 statistics in the zeroth term approximation (the phenomenological Tsallis distribution) is calculated in Eq.~(\ref{68}). This distribution corresponds to $\Lambda=0$, $\theta=1$ and the entropy $S=0$ for all values of temperature $T$ and volume $V$. The transverse momentum distribution (\ref{68}) in the rapidity range $y_{0}\leqslant y \leqslant y_{1}$ can be rewritten as
\begin{equation}\label{76}
\left. \frac{d^{2}N}{dp_{T}dy}\right|_{y_{0}}^{y_{1}} = \frac{gV}{(2\pi)^{2}} p_{T}  m_{T} \int\limits_{y_{0}}^{y_{1}} dy \cosh y  \left[1-(1-q) \frac{m_{T} \cosh y-\mu}{T} \right]^{\frac{q}{1-q}}.
\end{equation}

Let us apply the exact transverse momentum distribution of the Tsallis-3 statistics and the phenomenological Tsallis distribution (the transverse momentum distribution of the Tsallis-3 statistics in the zeroth therm approximation) to describe the experimental spectra of hadrons and compare their numerical results. The fits are performed using the statistical and systematic uncertainties of data. Figure~\ref{fig1} represents the transverse momentum spectra of $\pi^{-}$ mesons measured by the NA61/SHINE Collaboration~\cite{NA61} in $pp$ collisions at $\sqrt{s}=6.3,7.7,8.8,12.3$ and $17.3$ GeV in the rapidity range $0<y<0.2$. The solid and dashed curves are the fits of the experimental data to the exact Tsallis-3 distribution (\ref{75}) and the phenomenological Tsallis distribution (\ref{76}) integrated in the rapidity range $0<y<0.2$. The values of the fitting parameters of the exact Tsallis-3 distribution (\ref{75}) and the phenomenological Tsallis distribution (\ref{76}) are summarized in Table~\ref{t1} and  Table~\ref{t2}, respectively. The curves of these functions practically coincide and provide a good description of the experimental data. However, the values of the fitting parameters differ (see Table~\ref{t1} and Table~\ref{t2}). Thus, the phenomenological Tsallis distribution does not approximate well the exact transverse momentum distribution of the Tsallis-3 statistics. At NA61/SHINE energies, the temperature $T$ of the negatively charged pions in the Tsallis-3 statistics is approximately $T\sim 131$ MeV. However, the temperature $T$ of the phenomenological Tsallis distribution is lower than that of the Tsallis-3 statistics and it is approximately $T\sim 93$ MeV. The temperature of the phenomenological Tsallis distribution underestimates the temperature of the Tsallis-3 statistics. The radius $R$ of the system for the Tsallis-3 statistics is approximately $R\sim 5$ fm and the radius $R$ of the system for the phenomenological Tsallis distribution is approximately $R\sim 4.9$ fm. At the energies of the NA61/SHINE Collaboration, the values of the parameter $q$ for the Tsallis-3 statistics is approximately $q\sim 1.03$ and for the phenomenological Tsallis distribution is approximately $q\sim 1.06$. The parameter $q$ of the phenomenological Tsallis distribution overestimates the parameter $q$ of the Tsallis-3 statistics. These values of the parameter $q$ differ and they are close to unity. Note that the value of $\nu$ in Eq.~(\ref{71}) is chosen to provide the best fit of the experimental data. Note that the results for $\pi^{-}$ at NA61/SHINE energies obtained in this work by the phenomenological Tsallis distribution (the Tsallis-3 distribution in the zeroth term approximation) are practically the same as in Ref.~\cite{Parvan17a}.

\fulltable{\label{t1} Parameters of the Tsallis-3 statistics fit for the pions produced in $pp$ collisions at different energies. The chemical potential $\mu=0$.}
\begin{tabular}{@{}cccccccc}
 \br
 Collaboration  & Type        & $\sqrt{s}$, GeV &  $T$, MeV        &  $R$, fm          &  $q$                        &  $\chi^{2}/ndf$ & $\nu$  \\
 \mr
 NA61/SHINE     & $\pi^{-}$         & 6.3       & 125.45$\pm$4.23  & 4.502$\pm$0.171   & 1.0217$\pm$ $2 \cdot 10^{-6}$    & 2.70/15         & 0.4  \\
 NA61/SHINE     & $\pi^{-}$         & 7.7       & 127.45$\pm$12.49 & 4.893$\pm$0.589   & 1.0258$\pm$0.0053           & 1.15/15         & 0.4 \\
 NA61/SHINE     & $\pi^{-}$         & 8.8       & 131.20$\pm$7.66  & 4.864$\pm$0.371   & 1.0249$\pm$0.0039           & 0.82/15         & 0.4 \\
 NA61/SHINE     & $\pi^{-}$         & 12.3      & 133.02$\pm$6.47  & 5.330$\pm$0.342   & 1.0396$\pm$0.0050           & 0.77/15         & 0.4 \\
 NA61/SHINE     & $\pi^{-}$         & 17.3      & 136.13$\pm$5.74  & 5.515$\pm$0.310   & 1.0398$\pm$0.0044           & 0.44/15         & 0.4 \\
 PHENIX         & $\pi^{+}$         & 62.4      & 115.80$\pm$7.97  & 4.260$\pm$0.566   & 1.0694$\pm$0.0045           & 1.96/23         & 0.4 \\
 PHENIX         & $\pi^{-}$         & 62.4      & 111.44$\pm$10.86 & 4.563$\pm$0.813  & 1.0705$\pm$0.0062           & 1.11/23         & 0.4 \\
 PHENIX         & $\pi^{+}$         & 200       & 89.65$\pm$9.37   & 5.349$\pm$0.874   & 1.0913$\pm$0.0038           & 1.40/24         & 0.4 \\
 PHENIX         & $\pi^{-}$         & 200       & 100.05$\pm$9.28  & 4.610$\pm$0.700   & 1.0859$\pm$0.0040           & 0.92/24         & 0.4 \\
 ALICE          & $\pi^{+}$         & 900       & 91.74$\pm$2.14   & 5.093$\pm$0.103   & 1.0995$\pm$0.0016           & 3.19/30         & 0.4 \\
 ALICE          & $\pi^{-}$         & 900       & 94.22$\pm$3.57   & 4.960$\pm$0.165   & 1.0976$\pm$0.0024           & 1.38/30         & 0.4 \\
 ALICE          & $\pi^{+}+\pi^{-}$ & 2760      & 99.71$\pm$2.59   & 4.965$\pm$0.111   & 1.0782$\pm$0.0004           & 10.52/60        & 0.6 \\
 ALICE          & $\pi^{+}+\pi^{-}$ & 7000      & 96.44$\pm$1.38   & 6.219$\pm$0.074   & 1.1163$\pm$0.0006           & 11.45/38        & 0.6 \\
\br
\end{tabular}
\endfulltable

\fulltable{\label{t2} Parameters of the phenomenological Tsallis distribution (the Tsallis-3 distribution in the zeroth term approximation) fit for the pions produced in $pp$ collisions at different energies (the parameters for $\pi^{-}$ are practically the same as in Ref.~\cite{Parvan17a}). The chemical potential $\mu=0$.  }
\begin{tabular}{@{}cccccccc}
 \br
 Collaboration  & Type        & $\sqrt{s}$, GeV &  $T$, MeV        &  $R$, fm          &  $q$                        &  $\chi^{2}/ndf$   \\
 \mr
 NA61/SHINE     & $\pi^{-}$         & 6.3       & 96.76$\pm$8.69   & 4.431$\pm$0.344   & 1.0449$\pm$0.0223           & 2.70/15            \\
 NA61/SHINE     & $\pi^{-}$         & 7.7       & 92.68$\pm$7.67   & 4.782$\pm$0.334   & 1.0647$\pm$0.0208           & 1.14/15          \\
 NA61/SHINE     & $\pi^{-}$         & 8.8       & 95.39$\pm$7.33   & 4.749$\pm$0.301   & 1.0580$\pm$0.0204           & 0.99/15          \\
 NA61/SHINE     & $\pi^{-}$         & 12.3      & 91.04$\pm$7.44   & 5.172$\pm$0.350   & 1.0741$\pm$0.0209           & 0.89/15          \\
 NA61/SHINE     & $\pi^{-}$         & 17.3      & 91.18$\pm$7.57   & 5.357$\pm$0.376   & 1.0736$\pm$0.0205           & 0.46/15          \\
 PHENIX         & $\pi^{+}$         & 62.4      & 93.57$\pm$10.91  & 4.098$\pm$0.692   & 1.0893$\pm$0.0101           & 2.02/23          \\
 PHENIX         & $\pi^{-}$         & 62.4      & 89.67$\pm$10.69  & 4.370$\pm$0.769   & 1.0908$\pm$0.0098           & 1.17/23          \\
 PHENIX         & $\pi^{+}$         & 200       & 75.17$\pm$10.62  & 4.773$\pm$0.994   & 1.1266$\pm$0.0092           & 1.45/24          \\
 PHENIX         & $\pi^{-}$         & 200       & 82.81$\pm$10.37  & 4.234$\pm$0.772   & 1.1174$\pm$0.0089           & 1.01/24          \\
 ALICE          & $\pi^{+}$         & 900       & 73.14$\pm$2.36   & 4.778$\pm$0.129   & 1.1482$\pm$0.0051           & 4.08/30          \\
 ALICE          & $\pi^{-}$         & 900       & 74.68$\pm$2.37   & 4.686$\pm$0.125   & 1.1446$\pm$0.0051           & 2.18/30          \\
 ALICE          & $\pi^{+}+\pi^{-}$ & 2760      & 83.84$\pm$1.61   & 4.021$\pm$0.084   & 1.1480$\pm$0.0017           & 18.74/60        \\
 ALICE          & $\pi^{+}+\pi^{-}$ & 7000      & 68.86$\pm$2.14   & 5.597$\pm$0.165   & 1.1784$\pm$0.0035           & 14.05/38         \\
\br
\end{tabular}
\endfulltable

\begin{figure}[!htb]
\begin{center}
\includegraphics[width=0.8\textwidth]{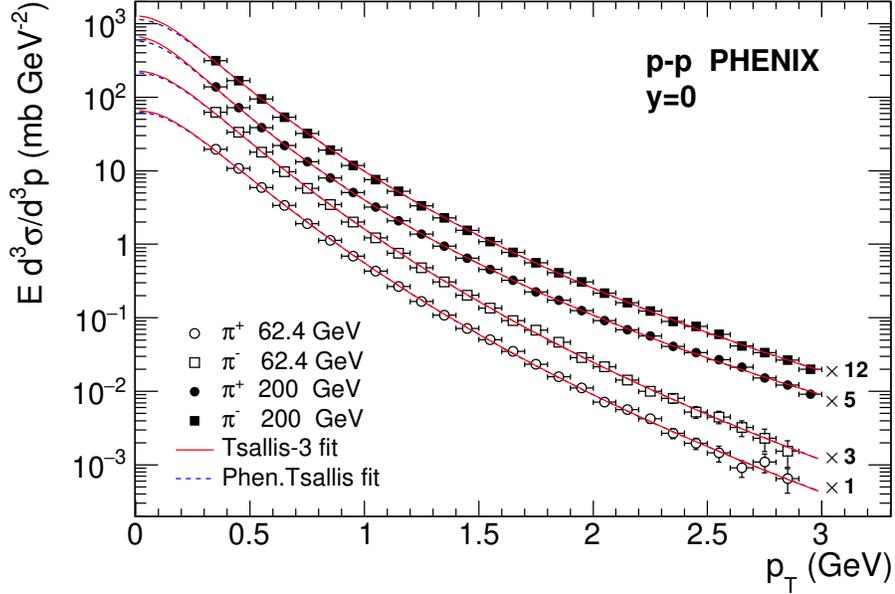} \vspace{-0.3cm}
\caption{(Color online) Transverse momentum spectra of $\pi^{-}$ and $\pi^{+}$ mesons measured by the PHENIX Collaboration~\cite{PHENIX} in $pp$ collisions at $\sqrt{s}=200$ and $62.4$ GeV at midrapidity. The solid and dashed curves are the fits of the data to the exact Tsallis-3 distribution and the phenomenological Tsallis distribution (the Tsallis-3 distribution in the zeroth term approximation), respectively, at rapidity $y=0$. The symbols represent the experimental data. The numbers next to the lines denote the scaling factor.} \label{fig2}
\end{center}
\end{figure}

Figure~\ref{fig2} represents the transverse momentum spectra of $\pi^{-}$ and $\pi^{+}$ mesons measured by the PHENIX Collaboration~\cite{PHENIX} in the proton-proton collisions at $\sqrt{s}=200$ and $62.4$ GeV at midrapidity. The solid curves are the fits of the experimental data to the exact Tsallis-3 distribution (\ref{74}) divided by the geometrical factor $2\pi p_T$:
\begin{eqnarray}\label{77}
 \frac{1}{2\pi p_{T}} \frac{d^{2}N}{dp_{T}dy} &=&  \frac{gV}{(2\pi)^{3}}  m_{T} \cosh y \frac{1}{\theta} \sum\limits_{n=0}^{n_{0}} \frac{\omega^{n}}{n!\Gamma\left(\frac{q}{q-1}\right)}  \int\limits_{0}^{\infty} t^{\frac{1}{q-1}-n} \nonumber \\ &\times&  e^{-t+\beta'(\Lambda-m_{T} \cosh y+\mu(n+1))}   (K_{2}\left(\beta' m \right) )^{n} dt.
\end{eqnarray}
The dashed curves are the fits of the experimental data to the phenomenological Tsallis distribution (\ref{68}) divided by the geometrical factor $2\pi p_T$. These distributions are multiplied by the total proton-proton cross section, which is $13.7\pm 1.5$ mb for $\sqrt{s}=62.4$ GeV and $23\pm 2.2$ mb for $\sqrt{s}=200$ GeV~\cite{PHENIX} (see~\ref{App1}). The theoretical curves have been calculated for $y=0$. The values of the parameters of the exact Tsallis-3 distribution (\ref{77}) are given in Table~\ref{t1} and the values of the parameters for the phenomenological Tsallis distribution (\ref{68}) divided by the geometrical factor $2\pi p_T$ are summarized in Table~\ref{t2}. The experimental data are fitted well by the curves of both the Tsallis-3 statistics and the phenomenological Tsallis distribution. However, the values of the fitting parameters for the phenomenological Tsallis distribution do not coincide with the values of the parameters for the exact Tsallis-3 distribution. Thus, at PHENIX energies the phenomenological Tsallis distribution does not approximate the exact Tsallis-3 distribution well. At the energy $\sqrt{s}=62.4$ GeV, the temperature $T$ of the Tsallis-3 statistics is approximately $T\sim 114$ MeV and the temperature $T$ of the phenomenological Tsallis distribution is lower than that of the Tsallis-3 statistics and it is approximately $T\sim 92$ MeV. The radius $R$ of the system for the Tsallis-3 statistics at $\sqrt{s}=62.4$ GeV is approximately $R\sim 4.4$ fm and the radius $R$ of the system for the phenomenological Tsallis distribution is approximately $R\sim 4.2$ fm. At $\sqrt{s}=62.4$ GeV, the value of the parameter $q$ for the Tsallis-3 statistics is approximately $q\sim 1.07$ and for the phenomenological Tsallis distribution it is approximately $q\sim 1.09$. At the energy $\sqrt{s}=200$ GeV, the temperature $T$ of the Tsallis-3 statistics is approximately $T\sim 95$ MeV and the temperature $T$ of the phenomenological Tsallis distribution is lower than that of the Tsallis-3 statistics and it is approximately $T\sim 79$ MeV. The radius $R$ of the system for the Tsallis-3 statistics at $\sqrt{s}=200$ GeV is approximately $R\sim 4.98$ fm and the radius $R$ of the system for the phenomenological Tsallis distribution is approximately $R\sim 4.5$ fm. At $\sqrt{s}=200$ GeV, the value of the parameter $q$ for the Tsallis-3 statistics is approximately $q\sim 1.09$ and for the phenomenological Tsallis distribution it is higher than that of the Tsallis-3 statistics and is approximately $q\sim 1.12$. Thus, at PHENIX energies, the temperature of the phenomenological Tsallis distribution underestimates the temperature of the Tsallis-3 statistics and the parameter $q$ of the phenomenological Tsallis distribution overestimates the parameter $q$ of the Tsallis-3 statistics. Note that the results for $\pi^{-}$ at PHENIX energies obtained in this work by the phenomenological Tsallis distribution (the Tsallis-3 distribution in the zeroth term approximation) are practically the same as in Ref.~\cite{Parvan17a}.

\begin{figure}[!htb]
\begin{center}
\includegraphics[width=0.8\textwidth]{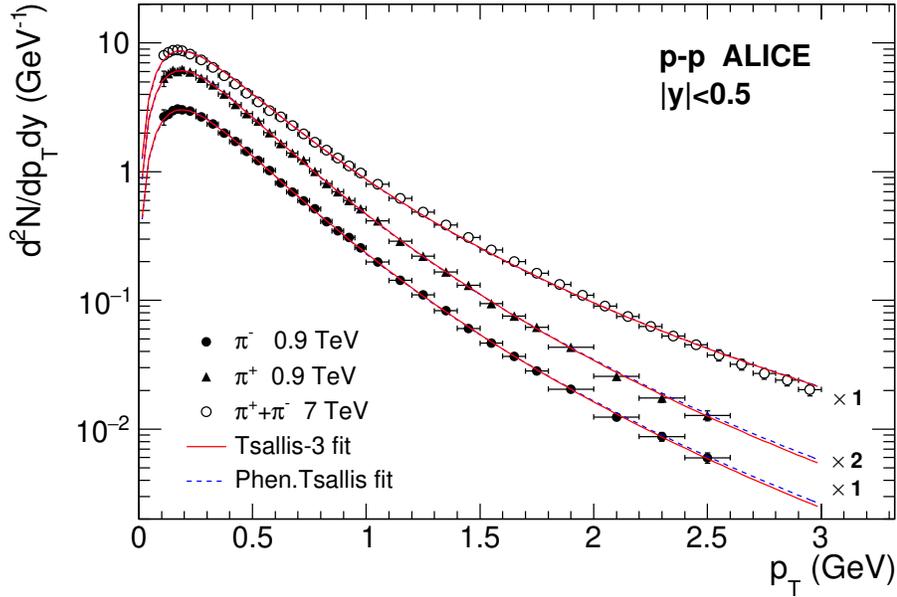} \vspace{-0.3cm}
\caption{(Color online) Transverse momentum spectra of $\pi^{-}$, $\pi^{+}$ and $\pi^{+}+\pi^{-}$ mesons measured by the ALICE Collaboration in $pp$ collisions at $\sqrt{s}=0.9$ TeV~\cite{ALICE09} and $\sqrt{s}=7$ TeV~\cite{ALICE7} in the rapidity interval $|y|<0.5$. The solid and dashed curves are the fits of the data to the exact Tsallis-3 distribution (\ref{75}) and the phenomenological Tsallis distribution (the Tsallis-3 distribution in the zeroth term approximation (\ref{76})), respectively. The symbols represent the experimental data. The numbers next to the lines denote the scaling factor.} \label{fig3}
\end{center}
\end{figure}

Figure~\ref{fig3} represents the transverse momentum spectra of $\pi^{-}$, $\pi^{+}$ and $\pi^{+}+\pi^{-}$ mesons measured by the ALICE Collaboration in the $pp$ collisions at $\sqrt{s}=0.9$ TeV~\cite{ALICE09} and $7$ TeV~\cite{ALICE7} in the rapidity range $|y|<0.5$. The solid and dashed curves are the fits of the experimental data to the exact Tsallis-3 distribution (\ref{75}) and the phenomenological Tsallis distribution (\ref{76}) integrated in the rapidity range $|y|<0.5$. The values of the fitting parameters of the exact Tsallis-3 distribution (\ref{75}) and the phenomenological Tsallis distribution (\ref{76}) are summarized in Table~\ref{t1} and  Table~\ref{t2}, respectively. The curves of these functions practically coincide with each other. They give a good description of the experimental data. However, the values of the fitting parameters of the phenomenological Tsallis distribution do not coincide with the values of the parameters of the exact Tsallis-3 distribution. Thus, at ALICE energies ($\sqrt{s}=0.9$ and $7$ TeV), the phenomenological Tsallis distribution does not approximate the exact Tsallis-3 distribution well. At the energy $\sqrt{s}=0.9$ TeV, the temperature $T$ of the Tsallis-3 statistics is approximately $T\sim 93$ MeV. The temperature $T$ of the phenomenological Tsallis distribution is lower than that of the Tsallis-3 statistics and it is approximately $T\sim 74$ MeV. The radius $R$ of the system for the Tsallis-3 statistics at $\sqrt{s}=0.9$ TeV is approximately $R\sim 5$ fm and the radius $R$ of the system for the phenomenological Tsallis distribution is approximately $R\sim 4.7$ fm. At $\sqrt{s}=0.9$ TeV, the value of the parameter $q$ for the Tsallis-3 statistics is approximately $q\sim 1.10$ and for the phenomenological Tsallis distribution it is approximately $q\sim 1.15$. At the energy $\sqrt{s}=7$ TeV, the temperature $T$ of $\pi^{+}+\pi^{-}$ pions for the Tsallis-3 statistics is approximately $T\sim 96$ MeV and the temperature $T$ for the phenomenological Tsallis distribution is lower than that of the Tsallis-3 statistics and it is approximately $T\sim 69$ MeV. The radius $R$ of the system for the Tsallis-3 statistics at $\sqrt{s}=7$ TeV is approximately $R\sim 6.2$ fm and the radius $R$ of the system for the phenomenological Tsallis distribution is approximately $R\sim 5.6$ fm. At $\sqrt{s}=7$ TeV, the value of the parameter $q$ for the Tsallis-3 statistics is approximately $q\sim 1.12$ and for the phenomenological Tsallis distribution it is approximately $q\sim 1.18$. Thus, at ALICE energies ($\sqrt{s}=0.9$ and $7$ TeV), the temperature of the phenomenological Tsallis distribution underestimates the temperature of the Tsallis-3 statistics and the parameter $q$ of the phenomenological Tsallis distribution overestimates the parameter $q$ of the Tsallis-3 statistics. Note that the results for $\pi^{-}$ at the ALICE energy $\sqrt{s}=0.9$ TeV obtained in this work by the phenomenological Tsallis distribution (the Tsallis-3 distribution in the zeroth term approximation) are practically the same as in Ref.~\cite{Parvan17a}.

\begin{figure}[!htb]
\begin{center}
\includegraphics[width=0.8\textwidth]{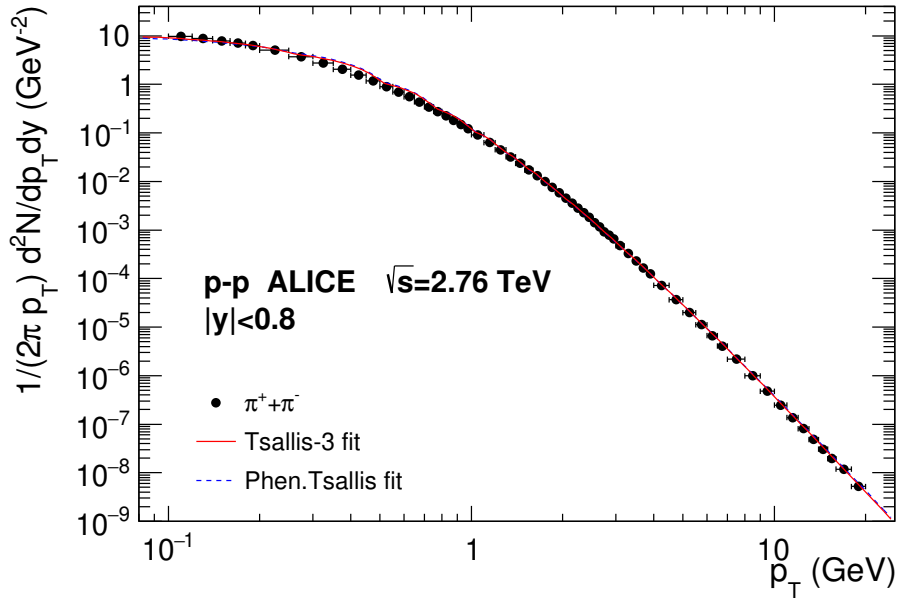} \vspace{-0.3cm}
\caption{(Color online) Transverse momentum spectra of $\pi^{+}+\pi^{-}$ mesons measured by the ALICE Collaboration~\cite{ALICE2.76} in $pp$ collisions at $\sqrt{s}=2.76$ TeV in the rapidity range $|y|<0.8$. The solid and dashed curves are the fits of the data to the exact Tsallis-3 distribution and the phenomenological Tsallis distribution (the Tsallis-3 distribution in the zeroth term approximation), respectively, divided by the geometrical factor $2\pi p_T$. The symbols represent the experimental data.} \label{fig4}
\end{center}
\end{figure}

Figure~\ref{fig4} represents the transverse momentum spectra of $\pi^{+}+\pi^{-}$ mesons measured by the ALICE Collaboration in $pp$ collisions at $\sqrt{s}=2.76$ TeV~\cite{ALICE2.76} in the rapidity interval $|y|<0.8$. The solid and dashed curves are the fits of the experimental data to the exact Tsallis-3 distribution (\ref{75}) and the phenomenological Tsallis distribution (\ref{76}), respectively, divided by the geometrical factor $2\pi p_T$ and integrated in the rapidity range $|y|<0.8$. The values of the fitting parameters of the exact Tsallis-3 distribution and the phenomenological Tsallis distribution are summarized in Table~\ref{t1} and  Table~\ref{t2}, respectively. The curves of the exact Tsallis-3 distribution and the phenomenological Tsallis distribution practically coincide. They give a good description of the experimental data. However, the values of the fitting parameters of the phenomenological Tsallis distribution do not coincide with the values of the parameters of the exact Tsallis-3 transverse momentum distribution. Thus, at ALICE energy $\sqrt{s}=2.76$ TeV, the phenomenological Tsallis distribution does not approximate the exact Tsallis-3 distribution well. At the energy $\sqrt{s}=2.76$ TeV, the temperature $T$ of the Tsallis-3 statistics is approximately $T\sim 100$ MeV and the temperature $T$ of the phenomenological Tsallis distribution is lower than that of the Tsallis-3 statistics and it is approximately $T\sim 84$ MeV. The radius $R$ of the system for the Tsallis-3 statistics at $\sqrt{s}=2.76$ TeV is approximately $R\sim 5$ fm and the radius $R$ of the system for the phenomenological Tsallis distribution is approximately $R\sim 4$ fm. At $\sqrt{s}=2.76$ TeV, the value of the parameter $q$ for the Tsallis-3 statistics is approximately $q\sim 1.08$ and for the phenomenological Tsallis distribution it is approximately $q\sim 1.15$. Thus, at ALICE energy $\sqrt{s}=2.76$ TeV, the temperature of the phenomenological Tsallis distribution underestimates the temperature of the Tsallis-3 statistics and the parameter $q$ of the phenomenological Tsallis distribution overestimates the parameter $q$ of the Tsallis-3 statistics.

\begin{figure}[!htb]
\begin{center}
\includegraphics[width=0.8\textwidth]{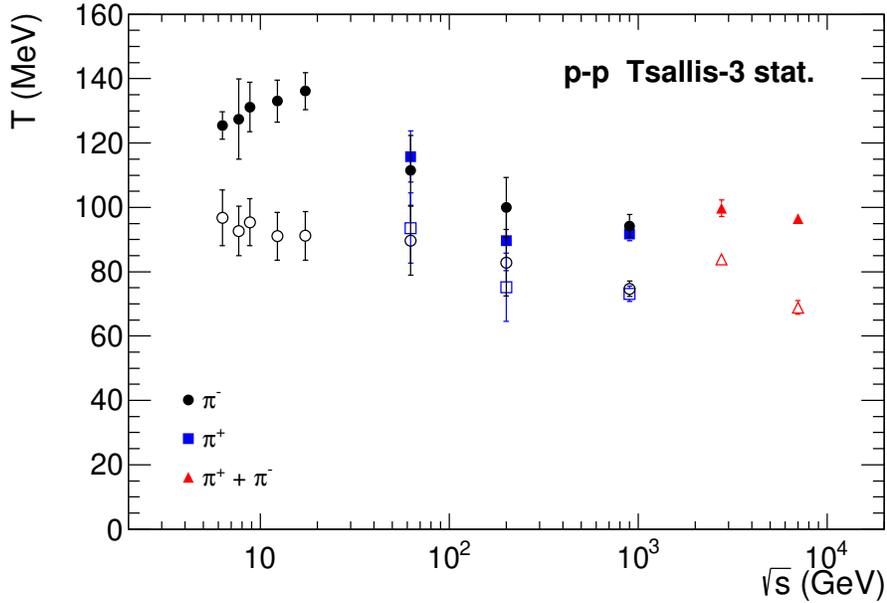} \vspace{-0.3cm}
\caption{(Color online) The energy dependence of the temperature $T$ of the Tsallis-3 statistics. The solid and open symbols are the results of the fit by the exact Tsallis-3 distribution and the phenomenological Tsallis distribution (the Tsallis-3 distribution in the zeroth term approximation), respectively, for the spectra of the $\pi^{-}$ (circle), $\pi^{+}$ (square) and $\pi^{+}+\pi^{-}$ (triangle) mesons measured by the NA61/SHINE~\cite{NA61}, PHENIX~\cite{PHENIX} and ALICE~\cite{ALICE09,ALICE7,ALICE2.76} Collaborations in $pp$ collisions at different energies.}  \label{fig5}
\end{center}
\end{figure}

Figure~\ref{fig5} represents the energy dependence of the temperature $T$ of the exact Tsallis-3 distribution and the phenomenological Tsallis distribution for $\pi^{-}$, $\pi^{+}$ and $\pi^{+}+\pi^{-}$ mesons produced in $pp$ collisions in the energy range $6.3$ GeV $\leqslant \sqrt{s}\leqslant 7$ TeV. The values of the temperature $T$ for the exact Tsallis-3 distribution and the phenomenological Tsallis distribution are compared. The solid and open points are the results of the fit for the Tsallis-3 statistics and the phenomenological Tsallis distribution (the transverse momentum distribution of the Tsallis-3 statistics in the zeroth term approximation), respectively. It is clearly seen that the temperature $T$ of the phenomenological Tsallis distribution differs essentially from the temperature of the Tsallis-3 statistics in the whole energy range. The temperature of the phenomenological Tsallis distribution underestimates the temperature of the Tsallis-3 statistics. Thus, the temperature $T$ of the phenomenological Tsallis distribution does not approximate well the temperature $T$ of the Tsallis-3 statistics distribution. Note that both temperatures of the pions decrease with the collision energy and have the same trend behavior.

\begin{figure}[!htb]
\begin{center}
\includegraphics[width=0.8\textwidth]{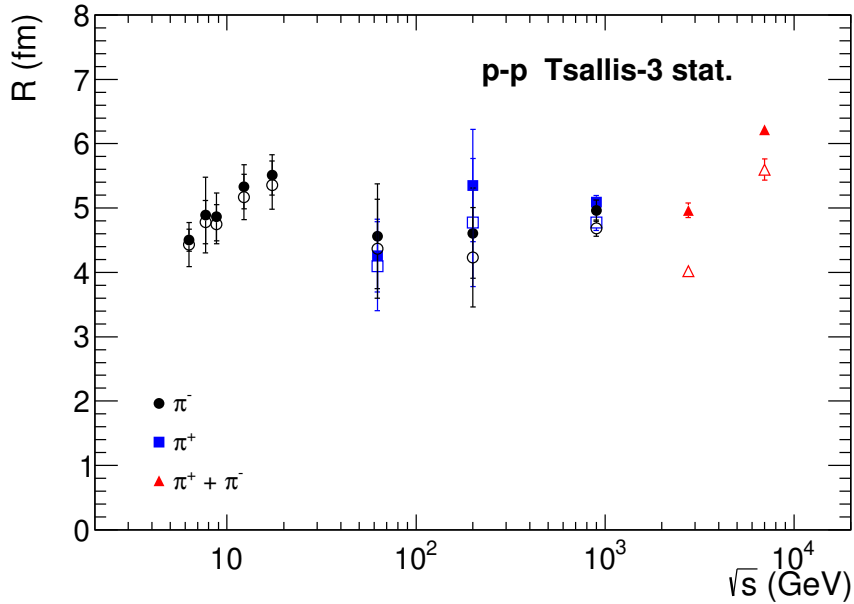} \vspace{-0.3cm}
\caption{(Color online) The energy dependence of the radius $R$ for the Tsallis-3 statistics. The notations are the same as in Fig.~\ref{fig5}.} \label{fig6}
\end{center}
\end{figure}

Figure~\ref{fig6} represents the energy dependence of the radius $R$ of the system for the Tsallis-3 statistics and the phenomenological Tsallis distribution for $\pi^{-}$, $\pi^{+}$ and $\pi^{+}+\pi^{-}$ mesons produced in $pp$ collisions in the energy range $6.3$ GeV $\leqslant \sqrt{s}\leqslant 7$ TeV. The values of the radius $R$ of the system for the Tsallis-3 statistics and the phenomenological Tsallis distribution are compared. The solid and open points are the results of the fit for the Tsallis-3 statistics and the phenomenological Tsallis distribution (the transverse momentum distribution of the Tsallis-3 statistics in the zeroth term approximation), respectively. The values of the radius $R$ of the system for the phenomenological Tsallis distribution do not coincide with the values of the radius $R$ of the system for the Tsallis-3 statistics. This difference increases with the energy of collision. However, at the energy up to $0.9$ TeV, the values of the radius $R$ for both model functions are compatible within the error uncertainties. Thus, the parameter $R$ of the phenomenological Tsallis distribution does not approximate well the parameter $R$ of the distribution of the Tsallis-3 statistics only at high collision energies. Note that the radii $R$ of the systems for both statistical distributions of pions are practically independent of the energy of collision and show the same trend behavior.

\begin{figure}[!htb]
\begin{center}
\includegraphics[width=0.8\textwidth]{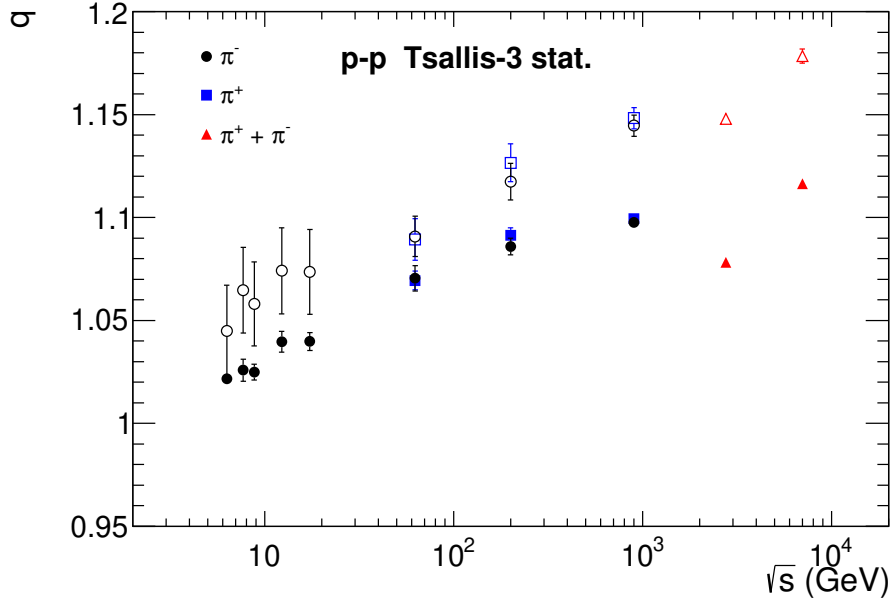} \vspace{-0.3cm}
\caption{(Color online) The energy dependence of the entropic parameter $q$ of the Tsallis-3 statistics. The notations are the same as in Fig.~\ref{fig5}.} \label{fig7}
\end{center}
\end{figure}

Figure~\ref{fig7} represents the energy dependence of the parameter $q$ for the Tsallis-3 statistics and the phenomenological Tsallis distribution for $\pi^{-}$, $\pi^{+}$ and $\pi^{+}+\pi^{-}$ mesons produced in $pp$ collisions in the energy range $6.3$ GeV $\leqslant \sqrt{s}\leqslant 7$ TeV. The values of the parameter $q$ for the Tsallis-3 statistics and the phenomenological Tsallis distribution are compared. The solid and open points are the results of the fit for the Tsallis-3 statistics and the phenomenological Tsallis distribution (the transverse momentum distribution of the Tsallis-3 statistics in the zeroth term approximation), respectively. It is clearly seen that the parameter $q$ of the phenomenological Tsallis distribution differs from the parameter $q$ of the Tsallis-3 statistics in the whole energy range. The difference between the values of these two parameters $q$ increases with energy. The parameter $q$ of the phenomenological Tsallis distribution overestimates the parameter $q$ of the Tsallis-3 statistics. Thus, the parameter $q$ of the phenomenological Tsallis distribution does not approximate well the parameter $q$ of the distribution of the Tsallis-3 statistics. However, the parameters $q$ of both model functions show the same trend behavior. The value $q=1$ corresponds to the Boltzmann-Gibbs statistics. The values of the parameter $q$ for the Tsallis-3 statistics and the phenomenological Tsallis distribution are not equal to unity and they increase significantly with $\sqrt{s}$. The same deviation of the transverse momentum distribution of the Tsallis-3 statistics from the Boltzmann-Gibbs exponential distribution is achieved at lower values of the parameter $q$ than that of the phenomenological Tsallis distribution. Note that in Figures~\ref{fig5}, \ref{fig6} and \ref{fig7}, the results for $\pi^{-}$ (open circles) obtained by the phenomenological Tsallis distribution (the Tsallis-3 distribution in the zeroth term approximation) are practically the same as in Ref.~\cite{Parvan17a}.

Let us estimate quantitatively the difference between the results of the Tsallis-3 statistics and the phenomenological Tsallis distribution (the transverse momentum distribution of the Tsallis-3 statistics in the zeroth term approximation) for three values of the energy of collision. At $\sqrt{s}=17.3$ GeV, for $\pi^{-}$ the difference between the temperatures of the Tsallis-3 statistics and the phenomenological Tsallis distribution is $44.95\pm 9.50$ MeV. Thus, the phenomenological Tsallis distribution temperature is lower than the Tsallis-3 statistics prediction at the level of $4.7\sigma$. These data indicate that there is a difference between the two results. The difference between the parameters $q$ is $0.0338\pm 0.0210$, and the phenomenological Tsallis distribution parameter $q$ is higher than the Tsallis-3 statistics prediction at the level of $1.6\sigma$. The difference between the two results for the parameter $q$ is not statistically significant. At $\sqrt{s}=0.9$ TeV, for $\pi^{+}$ the difference between the temperatures of the Tsallis-3 statistics and the phenomenological Tsallis distribution is $18.60\pm 3.19$ MeV. Thus, the phenomenological Tsallis distribution temperature is lower than the Tsallis-3 statistics prediction at the level of $5.8\sigma$. The difference between the parameters $q$ is $0.0487\pm 0.0053$. Therefore, the phenomenological Tsallis distribution parameter $q$ is higher than the Tsallis-3 statistics prediction at the level of $9\sigma$. For $\pi^{+}+\pi^{-}$ at $\sqrt{s}=7$ TeV, the difference between the temperatures $T$ is $27.58\pm 2.55$ MeV and the phenomenological Tsallis distribution temperature is lower than the Tsallis-3 statistics prediction at the level of $10.8\sigma$. The difference between the parameters $q$ is $0.0621\pm 0.0036$. Therefore, the phenomenological Tsallis distribution parameter $q$ is higher than the Tsallis-3 statistics prediction at the level of $17.5\sigma$. At $\sqrt{s}=0.9$ and $7$ TeV, the two results for the temperature $T$ and the parameter $q$ are different.

We can conclude that the transverse momentum distribution of the Tsallis-3 statistics in the zeroth term approximation (the phenomenological Tsallis distribution) estimates unsatisfactorily the parameters $(T,q)$ of the exact transverse momentum distribution of the Tsallis-3 statistics. Note that the values of the parameters of both these models have the same trend behavior as functions of the center of mass scattering energy because the phenomenological Tsallis distribution is an approximation of the exact Tsallis-3 distribution. The same is a cause for the compatibility of their volumes (compare, for example, Eqs.~(\ref{75}) and (\ref{76})) as the volume usually serves to fix the normalization of the distribution~\cite{Cleymans13}. However, the differences in the values of the temperatures $T$ and the parameters $q$ of both these models indicate the limitations of the phenomenological Tsallis distribution as an approximation of the exact Tsallis-3 distribution.

\section{Conclusions}\label{sec5}
Let us summarize the results of this paper. We have calculated the exact analytical expressions for the transverse momentum distributions of hadrons in the framework of the Tsallis-3 statistics for the Fermi-Dirac, Bose-Einstein and Maxwell-Boltzmann statistics of particles in the grand canonical ensemble and have applied the Maxwell-Boltzmann transverse momentum distribution of the Tsallis-3 statistics to describe experimental spectra of hadrons produced in proton-proton collisions at high energies. In the present paper, the new general formalism for the Tsallis-3 statistics in the grand canonical ensemble has been introduced. It was shown that this formalism is equivalent to other formulations of the Tsallis-3 statistics. In the new formulation of the Tsallis-3 statistics, the probability distribution of microstates of the system has been calculated from the principle of thermodynamic equilibrium. It was shown that the probability distribution is a function of two norm functions, which are the solutions of the system of two norm equations. The exact analytical results for the probability distribution of microstates, norm equations and the statistical averages were expressed in a general form in both the integral representation and series expansion. In the present paper, the transverse momentum distributions of the Tsallis-3 statistics for the Fermi-Dirac, Bose-Einstein and Maxwell-Boltzmann statistics of particles have been derived analytically for the first time. The results were expressed in both the integral representation and series expansion. The terms of the series expansion for the Maxwell-Boltzmann statistics of particles were given explicitly in the form of functions of the modified Bessel function of the second kind. The Maxwell-Boltzmann transverse momentum distribution was obtained for both the relativistic massive particles and the massless particles in the ultrarelativistic approximation. In the case of ultrarelativistic particles, the integrants can be integrated and the results can be expressed through the analytical functions.

We have also calculated the analytical expressions for the transverse momentum distributions of hadrons for the Tsallis-3 statistics in the zeroth term approximation in the case of the Fermi-Dirac, Bose-Einstein and Maxwell-Boltzmann statistics of particles and in the factorization approximation of the zeroth term approximation for the Fermi-Dirac and Bose-Einstein statistics of particles. We found that the classical and quantum transverse momentum distributions in the zeroth term approximation and the quantum transverse momentum distributions in the factorization approximation of the zeroth term approximation are the same in the Tsallis-3, Tsallis-2 and $q$-dual statistics. In the zeroth term approximation of the Tsallis-3 statistics, the entropy of the system is zero for all values of the temperature, volume, chemical potential and entropic parameter $q$. We have found that the phenomenological Tsallis distribution for the Maxwell-Boltzmann statistics of particles, which is extensively used in high energy physics, exactly coincides with the Maxwell-Boltzmann transverse momentum distribution in the zeroth term approximation for the Tsallis-3 statistics. Thus, the classical phenomenological Tsallis distribution is an approximation distribution for the classical exact Tsallis-3 distribution and it corresponds to the unphysical condition of zero entropy of the system in the Tsallis-3 statistics. We have also showed that both the quantum phenomenological Tsallis distribution for the Fermi-Dirac and Bose-Einstein statistics of particles and the classical and quantum Tsallis-like distributions for the Fermi-Dirac, Bose-Einstein and Maxwell-Boltzmann statistics of particles do not correspond to either the exact or approximate (the zeroth term approximation and the factorization approximation of the zeroth term approximation) transverse momentum distributions of the Tsallis-3 statistics.

In the present paper, the exact Maxwell-Boltzmann distribution of the Tsallis-3 statistics for $q>1$ has been applied to analyze the experimental data on hadrons produced in proton-proton collisions at high energies. We found that the Tsallis-3 distribution for $q>1$ is divergent. To regularize it, we introduced in the series expansions the upper cut-off limit of summation. We compared the numerical results of the classical phenomenological Tsallis distribution (the classical transverse momentum distribution in the zeroth term approximation of the Tsallis-3 statistics) with the exact Tsallis-3 distribution and applied them to describe the experimental data on charged pions produced in $pp$ collisions at high energies. The fitting parameters of the exact Tsallis-3 distribution and the phenomenological Tsallis distribution were found. The values of the parameters of both these model functions have the same trend behavior as functions of collision energy in the whole energy range. The values of the volumes (radii $R$) for the exact Tsallis-3 distribution and the phenomenological Tsallis distribution are compatible within the error uncertainties in the energy range up to $0.9$ TeV. However, the values of the temperature $T$ and the parameter $q$ of the phenomenological Tsallis distribution deviate from the values of the temperature $T$ and the parameter $q$ of the exact Tsallis-3 distribution in the whole energy range. Thus, the classical phenomenological Tsallis distribution is an unsatisfactory approximation for the exact classical transverse momentum distribution of the Tsallis-3 statistics.

%\vskip0.2in

\ack This work was supported in part by the RSCF grant, N22-72-10028. The author acknowledges the support by the Romanian Ministry of Research, Innovation and Digitalization, through the Project PN 23 21 01 01/2023.

\vskip0.2in

\noindent\textbf{Data availability statement}

\vskip0.2in

\noindent All data that support the findings of this study are included within the article (and any supplementary files).

\appendix{}

\section{Lorentz-invariant transverse momentum distribution}\label{App1}
The mean number of particles in the system is defined as
\begin{equation}\label{a1}
  N=\sum\limits_{\mathbf{p},\sigma} \langle n_{\mathbf{p}\sigma}\rangle,
\end{equation}
where $\langle n_{\mathbf{p}\sigma}\rangle$ is the mean number of particles (the mean occupation numbers) with momentum $\mathbf{p}$ and the third projection of the spin $\sigma$. The momentum of a particle is quantized and it is written as
\begin{equation}\label{a2}
  p_{i}=\frac{2\pi}{L} k_{i}, \qquad k_{i}=0,\pm 1, \pm2, \ldots, \;\; L=V^{1/3}, \;\; i=1,2,3,
\end{equation}
where $V$ is the volume of the system. Then, we have
\begin{equation}\label{a3}
  \sum\limits_{\mathbf{p},\sigma} \ldots = \frac{V}{(2\pi)^{3}} \int d^{3} p  \sum\limits_{\sigma} \ldots.
\end{equation}
Substituting Eq.~(\ref{a3}) into Eq.~(\ref{a1}), we obtain
\begin{equation}\label{a4}
  N=\frac{V}{(2\pi)^{3}} \int d^{3} p  \sum\limits_{\sigma} \langle n_{\mathbf{p}\sigma}\rangle = \int \frac{d^{3} p}{\varepsilon_{\mathbf{p}}} \left(\varepsilon_{\mathbf{p}} \frac{d^{3} N}{d^{3} p} \right),
\end{equation}
where
\begin{equation}\label{a5}
  \varepsilon_{\mathbf{p}} \frac{d^{3} N}{d^{3} p} =  \frac{V}{(2\pi)^{3}} \varepsilon_{\mathbf{p}} \sum\limits_{\sigma} \langle n_{\mathbf{p}\sigma}\rangle
\end{equation}
is the relativistic invariant distribution and $\varepsilon_{\mathbf{p}}$ is the single-particle energy. It is clearly seen that the multiplier $A$ in Eq.~(\ref{i1}) is
\begin{equation}\label{a5a}
  A=\frac{V}{(2\pi)^{3}} \varepsilon_{\mathbf{p}}.
\end{equation}

Let us introduce the hyperbolic coordinates
\begin{equation}\label{a6}
  p^{\mu}=(\varepsilon_{\mathbf{p}},p_{x},p_{y},p_{z})=(m_{T}\coth y,p_{T}\cos \varphi,p_{T}\sin \varphi,m_{T}\sinh y)
\end{equation}
with the measure
\begin{equation}\label{a7}
  d^{3} p = dp_{x} dp_{y} dp_{z}= \varepsilon_{\mathbf{p}} p_{T} dp_{T} dy d\varphi,
\end{equation}
where $\varepsilon_{\mathbf{p}}=m_{T} \cosh y$, $p_{T},y$ and $\varphi$ are the transverse momentum, rapidity and the azimuthal angle, respectively, and $m_{T}=\sqrt{p_{T}^{2}+m^{2}}$. Substituting Eq.~(\ref{a7}) into Eq.~(\ref{a5}) and integrating the result on $\varphi$, we have
\begin{equation}\label{a8}
  \frac{d^{2}N}{dp_{T}dy} = \frac{V}{(2\pi)^{3}} \int\limits_{0}^{2\pi} d\varphi  p_{T} \varepsilon_{\mathbf{p}} \sum\limits_{\sigma} \langle n_{\mathbf{p}\sigma}\rangle.
\end{equation}
Note that the Lorentz-invariant transverse momentum distribution in different coordinates can be written as
\begin{equation}\label{a9}
  \varepsilon_{\mathbf{p}} \frac{d^{3} N}{d^{3} p} =  \frac{d^{3} N}{p_{T} dp_{T} dy d\varphi} =  \frac{d^{3} N}{m_{T} dm_{T} dy d\varphi}.
\end{equation}
The Lorentz-invariant differential cross-section for a particle production can be written as
\begin{equation}\label{a10}
  \varepsilon_{\mathbf{p}} \frac{d^{3} \sigma}{d^{3} p} = \sigma_{tot} \times \varepsilon_{\mathbf{p}} \frac{d^{3} N}{d^{3} p},
\end{equation}
where $\sigma_{tot}$ is the total cross-section for the $pp$ or $AA$ collisions.

\section{Representations of the Tsallis-3 statistics}\label{App2}
Let us find two equivalent representations for the Tsallis-3 statistics. Using Eqs.~(\ref{8}) and (\ref{9}), we obtain
\begin{equation}\label{1a}
 p_{i} = \frac{1}{\overline{Z}} \left[1-(1-q)\frac{E_{i} - \langle H \rangle - \mu (N_{i}-\langle N \rangle)}{T\theta}\right]^{\frac{1}{1-q}}
\end{equation}
and
\begin{equation}\label{2a}
  \overline{Z} = \sum\limits_{i} \left[1-(1-q)\frac{E_{i} - \langle H \rangle - \mu (N_{i}-\langle N \rangle)}{T\theta}\right]^{\frac{1}{1-q}},
\end{equation}
where
\begin{equation}\label{3a}
  \overline{Z}^{1-q} \equiv \theta.
\end{equation}
Compare Eqs.~(\ref{1a})--(\ref{3a}) with Eqs.~(23), (24) and (28) of the canonical ensemble of the Tsallis-3 statistics given in Ref.~\cite{Tsal98}. Note that Eq.~(\ref{3a}) can be derived from Eqs.~(\ref{2}), (\ref{1b}) and (\ref{1a}).

In the probability distribution (\ref{1a}) there are two unknown quantities $\theta$ and $\langle H \rangle - \mu \langle N \rangle$. Thus, we should solve two norm equations to fix the probability distribution. Substituting Eq.~(\ref{1a}) into Eq.~(\ref{3}) and using Eq.~(\ref{3a}), we obtain
\begin{equation}\label{4a}
  \overline{Z} = \sum\limits_{i} \left[1-(1-q)\frac{E_{i} - \langle H \rangle - \mu (N_{i}-\langle N \rangle)}{T\theta}\right]^{\frac{q}{1-q}}.
\end{equation}
Compare Eq.~(\ref{4a}) with Eq.~(9) of the canonical ensemble of the Tsallis-3 statistics given in Ref.~\cite{Abe2000}.
Equating Eqs.~(\ref{2a}) and (\ref{4a}), we find the first norm equation
\begin{eqnarray}\label{5a}
 && \sum\limits_{i} \left[1-(1-q)\frac{E_{i} - \langle H \rangle - \mu (N_{i}-\langle N \rangle)}{T\theta}\right]^{\frac{q}{1-q}}   \nonumber   \\
 && = \sum\limits_{i} \left[1-(1-q)\frac{E_{i} - \langle H \rangle - \mu (N_{i}-\langle N \rangle)}{T\theta}\right]^{\frac{1}{1-q}}.
\end{eqnarray}
The second norm equation can be written as
\begin{eqnarray}\label{6a}
  \langle H \rangle - \mu \langle N \rangle &=& \frac{1}{\overline{Z}}\sum\limits_{i} (E_{i} - \mu N_{i} ) \nonumber  \\
  && \left[1-(1-q)\frac{E_{i} - \langle H \rangle - \mu (N_{i}-\langle N \rangle)}{T\theta}\right]^{\frac{q}{1-q}}.
\end{eqnarray}
The solutions of the norm equations (\ref{5a}) and (\ref{6a}) are $\theta$ and $\langle H \rangle - \mu \langle N \rangle$. They fix the probability distribution (\ref{1a}) and the statistical averages, which can be written as
\begin{equation}\label{7a}
  \langle A \rangle = \frac{1}{\overline{Z}}\sum\limits_{i} A_{i}  \left[1-(1-q)\frac{E_{i} - \langle H \rangle - \mu (N_{i}-\langle N \rangle)}{T\theta}\right]^{\frac{q}{1-q}}.
\end{equation}
Note that the probability distribution (\ref{1a}) is equivalent to the probability distribution (\ref{10}).

Let us find another representation of the Tsallis-3 statistics. The probability distribution (\ref{1a}) can be rewritten as
\begin{equation}\label{8a}
 p_{i} = \frac{1}{Z'} \left[1-(1-q)\beta'_{*} (E_{i} - \mu N_{i})\right]^{\frac{1}{1-q}},
\end{equation}
where
\begin{equation}\label{9a}
 Z' \equiv \sum\limits_{i} \left[1-(1-q)\beta'_{*} (E_{i} - \mu N_{i})\right]^{\frac{1}{1-q}}
\end{equation}
and
\begin{equation}\label{10a}
  \beta'_{*} \equiv \frac{\beta}{\theta + (1-q) \beta (\langle H \rangle - \mu \langle N \rangle)}, \qquad \beta \equiv \frac{1}{T}.
\end{equation}
Compare Eqs.~(\ref{8a})--(\ref{10a}) with Eqs.~(39) and (40) of the canonical ensemble for the Tsallis-3 statistics given in Ref.~\cite{Tsal98}. Note that the probability distribution (\ref{8a}) is equivalent to the probability distributions (\ref{10}) and (\ref{1a}) in terms of the independent variables of the state $(T,V,\mu,q)$. Thus, all considered representations of the Tsallis-3 statistics are equivalent.

%\vskip0.2in

\end{document}